\documentclass[%
 aip,
 amsmath,amssymb,
 reprint,%
]{revtex4-2}
\usepackage{graphicx} % Required for inserting images
\usepackage{amsmath}
\usepackage{dcolumn}
\usepackage{bigints}
\usepackage{float}
\usepackage{color}
\usepackage{etoolbox}

\DeclareMathOperator\erfc{erfc}
\usepackage[utf8]{inputenc}
\usepackage[T1]{fontenc}

\makeatletter
\def\@email#1#2{%
 \endgroup
 \patchcmd{\titleblock@produce}
  {\frontmatter@RRAPformat}
  {\frontmatter@RRAPformat{\produce@RRAP{*#1\href{mailto:#2}{#2}}}\frontmatter@RRAPformat}
  {}{}
}%
\makeatother
\begin{document}

\preprint{AIP/123-QED}

%\title[]{Mean Force Emission Theory with a Pseudopotential for Studying Classical Bremsstrahlung in Electron-Ion Plasmas}
\title[]{Mean Force Emission Theory for Classical Bremsstrahlung in Electron-Ion Plasmas}
% Force line breaks with \\
\author{J.P. Kinney}
\author{H.J. LeFevre}%
% \email{Second.Author@institution.edu.}
%\affiliation{ 
%Authors' institution and/or address%\\This line break forced with \textbackslash\textbackslash
%}%

\author{C.C. Kuranz}
\author{S.D. Baalrud}
\email{baalrud@umich.edu}

%Department of Nuclear Engineering and Radiological Sciences, University of Michigan, Ann Arbor, MI 48109, USA
\affiliation{Department of Nuclear Engineering and Radiological Sciences, University of Michigan, Ann Arbor, MI 48109, USA}
 %\homepage{http://www.Second.institution.edu/~Charlie.Author.}
%\affiliation{
%Second institution and/or address%\\This line break forced% with \\
%}%

\date{\today}% It is always \today, today,

\begin{abstract}
This work extends the previously developed mean force emission theory to describe electron-ion plasmas. Results are compared to molecular dynamics simulations. 
%where 
The main extensions are to account for the attractive nature of electron-ion interactions and to model short-range quantum effects using the Kelbg potential. 
By reducing the electron-ion force inside the deBroglie wavelength, the Kelbg potential causes a decay at high frequencies and a decrease in magnitude of the low frequency bremsstrahlung spectrum. The attractive electron-ion interaction also allows for classically bound states that generate peaks in the emission spectrum. Results show that the Kelbg potential can capture quantum modifications to classical Gaunt factors, but is limited in describing emission at very high frequencies. This work further supports the notion that there is a peak in emission near the plasma frequency at strong coupling that cannot be captured using the common Drude correction. Importantly, the linear response framework used to calculate the bremsstrahlung emission coefficient is related to both the absorption coefficient and the real part of the dynamic electrical conductivity. This means that the conclusions drawn from this study can be applied to these transport coefficients as well. Finally, this work compares the results with commonly used classical and quantum mechanical Gaunt factors, and discusses the impact of a Fermi-Dirac distribution of electrons on emission and why screening slightly reduces the bremsstrahlung power in weakly coupled and non-degenerate plasmas.
\end{abstract}

\maketitle

\section{Introduction}
Bremsstrahlung emission and absorption have been recent topics of interest in the plasma physics community.~\cite{turnbull2023inverse,turnbull2024reconciling} They play an important role in radiation transport, including in dense, strongly correlated plasmas found in nuclear fusion experiments,~\cite{zylstra2022burning,sadler2019kinetic,chen2015application} astrophysical objects,~\cite{imamura1985on,mukai2017X,campana2002the,zhang2022imaging} and in extreme ultraviolet lithography used for semiconductor processing~\cite{nishihara2008plasma,campos2010the}. 
Both emission and absorption processes are associated with electron dynamics, which is notoriously difficult to describe in strongly correlated systems. Our recent work developed mean force emission theory to extend the description of classical bremsstrahlung into the strongly coupled regime.~\cite{kinney2024mean} Results of the model were compared with classical molecular dynamics (MD) simulations for a repulsively interacting (positron-ion) plasma. 
%and used the simulation results to benchmark mean force emission theory. 
Repulsive interactions were chosen to avoid Coulomb collapse in classical MD simulations, allowing a first-principles comparison between the model and simulations. 

Here, we extend this work to attractively interacting (electron-ion) plasmas. Coulomb collapse is avoided in the MD simulations by modeling short-range quantum effects through the Kelbg potential.~\cite{kelbg1963theorie,filinov2003improved} The quantum nature of the electron-ion interaction causes a decay in the  high frequency limit of the bremsstrahlung spectrum by reducing the electron-ion force inside the thermal deBroglie wavelength. Furthermore, the reduced force limits the scattering angle of electrons during close encounters with ions, leading to a decrease in the electron-ion collision frequency and a corresponding decrease in the low frequency bremsstrahlung emission. The results also show that there is a peak in emission near the plasma frequency at strong coupling that cannot be captured by the common method of correcting the low frequency limit with a Drude model,~\cite{johnson2006optical,starrett2016kubo,shaffer2017free} which has the same origin as with repulsive interactions.~\cite{kinney2024mean} Importantly, the general framework for classical bremsstrahlung emission used in this paper is related to both bremsstrahlung absorption and the dynamic conductivity in the linear response regime.~\cite{befeki1966radiation,dawson1962high,oster1961emission} Thus, the conclusions drawn here apply to these transport coefficients too and the same method can be used to calculate them.

This work is particularly motivated by radiation transport in warm dense matter.~\cite{haines2024charged,zastrau2008bremsstrahlung} Warm dense matter is a challenging regime for theory because electron-ion interactions are strongly coupled ($\Gamma = Zq^{2}/(4\pi\epsilon_{0}ak_\textrm{B}T) \geq 1$) and electrons are moderately degenerate ($\Theta = k_\textrm{B}T/E_\textrm{F} \sim 1$). Here, $a$ is the Wigner-Seitz radius, $q$ is the electron charge, $T$ is the temperature, $Z$ is the ion charge state, and $E_\textrm{F} = \hbar^{2}\left(3\pi^{2}n_{e}\right)^{2/3}/(2m_{e})$ is the Fermi energy of a three-dimensional free electron gas. 
Calculations of bremsstrahlung in this regime have been approached from both condensed matter and traditional plasma physics perspectives. 
On the condensed matter side, two classes of approaches are density functional theory molecular dynamics (DFT-MD)\cite{hu2014first} and average-atom models.~\cite{johnson2006optical,starrett2016kubo,shaffer2017free} These approaches have the benefit of rigorous inclusion of quantum-mechanical effects on scattering and electron degeneracy, but are more computationally intensive and harder to extend to higher temperature than approximate analytic models. 
Approximate analytic models\cite{landau1971classical,oster1961emission,befeki1966radiation,dawson1962high,sommerfeld1924atombau} typically offer quick formulas, with the downside of limited accuracy, particularly at strong coupling. The objective of this paper is to further develop the framework of mean force emission theory, with the eventual goal of providing fast and accurate calculations of bremsstrahlung emission across a wide range of plasma densities and temperatures, including the warm dense matter regime.

Here, the model is tested using classical MD simulations. 
One challenge in applying classical MD is that electrons and ions can get arbitrarily close to each other because of their attractive nature. 
This is not possible to resolve in a finite timestep, and leads to the phenomenon of Coulomb collapse.~\cite{hansen1978microscopic} 
Of course, Coulomb collapse is not physical, as quantum effects like the Heisenberg uncertainty and the Pauli exclusion principles that act at scales characteristic of the de~Broglie wavelength are responsible for the stability of matter.~\cite{lieb1976thestability} 
Unfortunately, it is not currently possible to do a first-principles simulation of the quantum many-body problem at these conditions, so some approximate method is necessary. 
Here, we take a pseudopotential approach, modeling the electron-ion interactions using the Kelbg potential. The Kelbg potential is a quantum statistical potential that was first derived from the two-particle density matrix in the high temperature limit,~\cite{kelbg1963theorie} and subsequently extended to lower temperatures.\cite{filinov2003improved} 
It has been applied to simulate dense plasmas~\cite{filinov2004temperature,graziani2012large} and is meant to capture the average two-particle quantum interaction at close distances. Qualitatively, the Kelbg potential becomes linear in the limit that the particle separation $r$ is much smaller than the thermal de Broglie wavelength $\lambda_\textrm{th} = \sqrt{2\pi\hbar^{2}/\left(m_{e}k_\textrm{B}T\right)}$. Even though for $\Theta \gg 1$ the plasma is said to be `classical', the rare close electron-ion collisions that occur at distances $r<\lambda_\textrm{th}$ are quantum-mechanical in nature. The Kelbg potential serves as an attempt to capture the average electron dynamics of these close interactions.

As expected, quantum effects at electron-ion separations less than the thermal de~Broglie wavelength are found to cause a rapid decay in the high frequency emission spectrum by limiting the force between an electron and ion. The limited force also decreases the electron-ion collision frequency, and thus reduces the magnitude of the low frequency emission spectrum. Importantly, as in the repulsive interaction, there is still a peak in emission that occurs at strong coupling due to correlated electron motion on intermediate timescales near the electron plasma period. Even with these modifications, the same framework for mean force emission theory is found to successfully describe the bremsstrahlung spectrum. 
A second unique feature of the MD simulations with an attractive interaction is that they allow for classically bound states to form between electrons and ions. These bound states are shown to impact the bremsstrahlung spectrum by introducing a peak that is associated with the orbit frequency of bound particles. A simple model that predicts the location of these peaks is discussed. 

Finally, using the Kelbg potential to calculate the bremsstrahlung spectrum helps to clarify analytic approaches and their attempt to include quantum-mechanical effects in the weakly coupled limit. Textbook approaches for classical attractively interacting systems predict a constant emission rate as the frequency $\omega\xrightarrow{}\infty$, which is not physical.~\cite{landau1971classical} Quantum modifications to the short range electron-ion interaction in the present work avoid this result. Standard approaches to model such quantum-mechanical effects apply ad-hoc modifications of the frequency dependent `Coulomb logarithm', or `Gaunt factor'.~\cite{befeki1966radiation,rybicki1979radiative} Mean force emission theory is shown to naturally capture these modifications. Although the Kelbg potential is a well-motivated approximation for incorporating aspects of the wave-like nature of interactions at short distances, it is not a complete quantum theory. For instance, the trajectories of particles are modeled entirely classically. Other work\cite{morozov2005molecular} has found the potential to have limitations in describing very high frequency phenomena as a consequence. This work supports this conclusion, and so the Kelbg potential should not be interpreted as an accurate representation of a physical warm dense matter system. Instead, it is used here to enable a first-principles test of a theory  using MD simulations for an attractively interacting system. We expect that validation of the mean force concept for this model system may inform later development of a fully quantum mechanical calculation.

This paper is organized as follows. Section~\ref{sec:MD} introduces the MD simulations and calculation of the bremsstrahlung emission coefficient. It also summarizes the qualitative influence of the attractive interactions and quantum effects on the emission spectra, the impact of bound states, and the relation between the bremsstrahlung emission and the force autocorrelation function. This correlation function relationship is used to show how the emission, absorption, and real part of dynamic conductivity are related in the linear response regime. Section~\ref{sec:theory} discusses mean force emission theory. Section~\ref{sec:discussion} compares mean force emission theory with analytic classical and quantum Gaunt factors at weak coupling, applies the theory to calculate the bremsstrahlung power density emitted from plasma, and, as a step towards a quantum-mechanical model, discusses how the exponential decay of the bremsstrahlung spectrum and a Fermi-Dirac distribution of electrons might be further included in the results. This section also discusses the limitation of the Kelbg potential in describing very high frequency emission. Section~\ref{sec:conclusion} discusses conclusions based on these results.

%\end{itemize}

\section{\label{sec:MD}Molecular Dynamics Simulations}

\subsection{\label{subsec:Setup} Setup}
%and Calculation of the Emission Coefficient (do not know how to shorten this)}

\begin{figure}[b!]\label{Kelbg_varyTheta}
\includegraphics[width=0.5\textwidth]{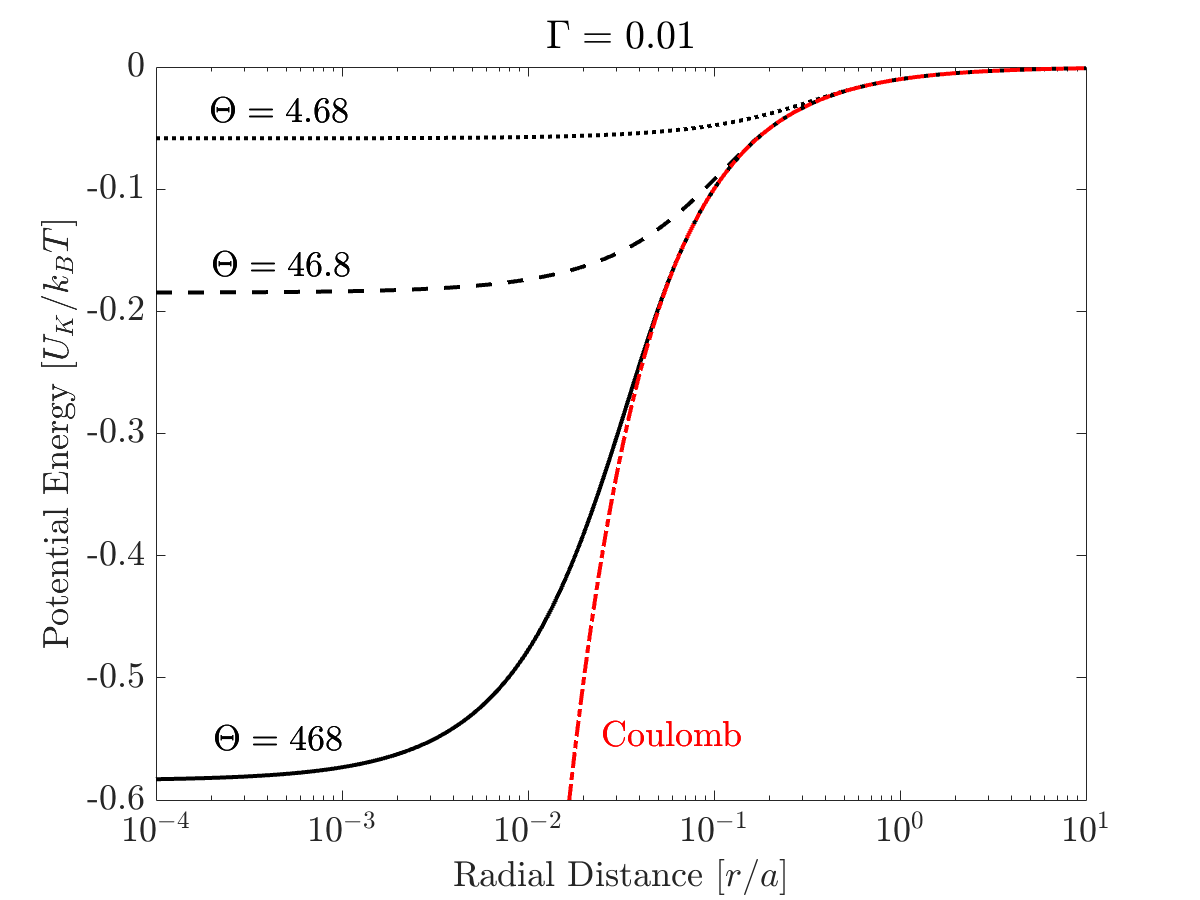}
\caption{\label{Kelbg_varyTheta} Normalized electron-ion Kelbg potentials for $\Gamma = 0.01$ and values of $\Theta = 468, 46.8,$ and $4.68$. The dashed-dotted red line indicates the normalized electron-ion Coulomb potential.}
\end{figure}

Molecular dynamics simulations were performed using LAMMPS (Large-scale Atomic/Molecular Massively Parallel Simulator).~\cite{thompson2022lammps} The simulations consisted of a two-component plasma of electrons and ions ($Z = 1$) at a fixed temperature ($T_\textrm{e} = T_\textrm{i} = T$) with a mass ratio of $m_\textrm{i}/m_\textrm{e}\approx1836$, corresponding to protons. Electron-electron and ion-ion interactions were modeled using the Coulomb potential $U_\textrm{C} = q^{2}/(4\pi\epsilon_{0} r)$. Because the attractive Coulomb potential brings electrons and ions arbitrarily close to each other, stability from Coulomb collapse requires that electron-ion interactions be modeled with a pseudopotential. In classical weakly coupled regimes, close collisions are sufficiently rare that the pseudopotential can be made to be purely a numerical stability requirement that does not influence the calculation results.~\cite{tiwari2017thermodynamic} Common pseudopotentials include repulsive\cite{tiwari2017thermodynamic,shaffer2019thebarkas} or soft core\cite{devriendt2022classical} potentials. As the plasma becomes more strongly coupled ($\Gamma\xrightarrow{}1$), or degenerate ($\Theta\xrightarrow{}1$), the form of the pseudopotential influences the calculated plasma properties. Since we are interested in parameter regimes ranging from strong to weak coupling and degeneracy, the pseudopotential is responsible for modeling the quantum effects associated with close interactions. %It is noted that even in a plasma with $\Theta > 1$, close electron-ion collisions ($r<\lambda_\textrm{th}$) are quantum-mechanical in nature. Thus, the electron-ion interaction at close distances needs to be described using quantum mechanics. 

Here, the Kelbg potential is used to model `average' properties of the quantum-mechanical interaction between an ion and electron at close distances. It has been used previously in classical MD simulations to study thermodynamical properties of dense hydrogen,~\cite{filinov2004temperature} temperature relaxation,~\cite{graziani2012large} optical conductivity,~\cite{morozov2005molecular} and the dynamic structure factor~\cite{golubnychiy2001dynamical}. It is derived from a high temperature expansion of the diagonal elements of the two-particle density matrix.~\cite{demyanov2022derivation} The Kelbg potential for $Z = 1$ is given in Ref.~\onlinecite{filinov2004temperature} as
\begin{eqnarray}
    \label{eq:KelbgPotential}
    U_\textrm{K}(r) = \frac{-q^{2}}{4\pi\epsilon_{0}r}\left[1-e^{-r^{2}/\lambda^{2}}
    +\sqrt{\pi}\frac{r}{\lambda\gamma}\erfc{\left(\gamma\frac{r}{\lambda}\right)}\right]
\end{eqnarray}
where $\lambda = \hbar/\sqrt{2m_\textrm{e}k_\textrm{B}T}$ and the parameter $\gamma$ is a temperature-dependent fit parameter that matches the analytic form of the Kelbg potential to the exact solution to the diagonal pair potential obtained from the two-particle density matrix using a matrix squaring technique.~\cite{filinov2004temperature} The fit parameter is given as 
\begin{eqnarray}
    \label{eq:GammaFit}
    \gamma(T) = \frac{x+x^{2}}{1+ax+x^{2}}
\end{eqnarray}
where $x = \sqrt{8\pi T/(315775)}$ and $a = 1.090$. In terms of the dimensionless parameters $\Gamma$ and $\Theta$,
 this is 
%the expression for $\Tilde{U}_\textrm{K} = U_\textrm{K}/(k_\textrm{B}T)$ for $Z = 1$ is
\begin{equation}
    \label{eq:KelbgPotentialDimensionless}
    \Tilde{U}_\textrm{K}(\Tilde{r}) = \frac{-\Gamma}{\Tilde{r}}\biggl[1-e^{-A^{2}\Tilde{r}^{2}\Theta}+\frac{A\sqrt{\pi\Theta}\Tilde{r}}{\gamma} \erfc{\left(A\gamma\sqrt{\Theta}\Tilde{r}\right)}\biggr]
    %&\times\left[1-\erf{\left(A\gamma\sqrt{\Theta}\Tilde{r}\right)}\right]\biggr)
\end{equation}
where $\Tilde{U}_\textrm{K} = U_\textrm{K}/(k_\textrm{B}T)$, $\Tilde{r} = r/a$ and $A = (9\pi/8)^{1/3}$. 

Figure~\ref{Kelbg_varyTheta} shows that the main qualitative characteristic of the Kelbg potential is that it is linear at short distances. The length scale at which this plateau occurs is associated with the parameter $\lambda$ which is simply related to the thermal deBroglie wavelength as $\lambda = [1/(2\sqrt{\pi})]\lambda_\textrm{th}$. As $\Theta$ decreases, $\lambda/a$ increases and the non-Coulombic behavior extends to larger radial distances. Consequently, for electron-ion separation distances $r \lesssim \lambda$, the force on the electron is approximately constant. The plateau at short distances is meant to capture the `wave-like' nature of the electron-ion interaction. More specifically, each electron with momentum $m_\textrm{e}v$ that interacts with an ion would have an associated de Broglie wavelength $\lambda_\textrm{dB}=h/\left(m_\textrm{e}v\right)$. The Kelbg potential attempts to capture the `average' electron by supposing that all electrons have the thermal de Broglie wavelength associated with the thermal speed $\lambda_\textrm{th}
\sim h/\left(m_\textrm{e}v_\textrm{Te}\right)$, with $v_\textrm{Te} = \sqrt{2k_{B}T/m_\textrm{e}}$.

For numerical efficiency, the MD simulations use the particle-particle particle-mesh (P3M) method.~\cite{frenkel2023understanding} In P3M, the short-range force on a particle is calculated directly within a specified distance ($r_\textrm{c}$), while outside of this distance charges are interpolated to a grid and the long-range force on a particle is advanced from the electric field computed at the grid cells. The long-range solver in this work remained Coulombic, so the Kelbg potential had to be properly input into the short range solver so as to minimize error in the P3M calculation. The short-range force implemented in LAMMPS for an interparticle separation $r$ is given as $\mathbf{F} = F(r)\hat{\mathbf{r}}$ where the magnitude is
\begin{eqnarray}
    \label{eq:forcemagnitudeLAMMPS}
    F(r) = -\left(\frac{dU_\textrm{K}}{dr}\right)\left[\erfc{\left(\alpha r\right)}+\frac{2\alpha r}{\sqrt{\pi}}\exp{\left(-\alpha^{2}r^{2}\right)}\right].
\end{eqnarray}
Here, $\alpha$ is the Ewald parameter that smoothly splits the force calculation between the short-range and long-range solvers.~\cite{frenkel2023understanding} This expression was chosen in order to match the force due to the Kelbg interaction and be consistent with the long range Coulombic solver. The values of the Ewald parameter $\alpha$ and the cutoff distance $r_\textrm{c}$ are given in Table~\ref{tab:table1} and were chosen in order to minimize the error in the force while making the calculations computationally feasible.

\begin{table}
\caption{\label{tab:table1} Values of the Ewald parameter $\alpha$ and cutoff distance $r_\textrm{c}$ used for the P3M calculation for various $\Gamma$ and $\Theta$.}
\begin{ruledtabular}
\begin{tabular}{cccc}
 $\Gamma$ &$\Theta$ &$\alpha$ [$m^{-1}$]
 &$r_c$\\
\hline
0.01& 468 & $3.16\times10^{9}$ & $5a$\\
0.01& 46.8 & $2.60\times10^{10}$ & $5a$\\
0.01& 4.68 & $2.60\times10^{11}$ & $5a$\\
0.1& 46.8 & $3.16\times10^{9}$ & $5a$\\
0.1& 4.68 & $2.60\times10^{10}$ & $5a$\\
0.1& 0.468 & $9.00\times10^{10}$ & $10a$\\
1& 4.68 & $3.16\times10^{9}$ & $5a$\\
1& 0.468 & $9.00\times10^{9}$ & $10a$\\
\end{tabular}
\end{ruledtabular}
\end{table}

Each simulation used a three dimensional box with periodic boundary conditions containing $10^{4}$ particles. Prior to data gathering, each plasma was equilibrated to a fixed $\Gamma$ and $\Theta$ value using a Nose-Hoover thermostat.~\cite{frenkel2023understanding} Once the particles settled into their equilibrium spatial and velocity distributions the thermostat was turned off, and the total energy in the simulation was fixed. In order to fully resolve all collisions and the high-frequency emission spectrum, the time step was chosen to be  $10^{-4}\omega_\textrm{pe}^{-1}$, where $\omega_\textrm{pe} = \sqrt{n_\textrm{e}q^{2}/(m_\textrm{e}\epsilon_{0})}$ is the electron plasma frequency. The total runtime for each simulation was $10^{3}\omega_\textrm{pe}^{-1}$.

Of course, the Kelbg potential is still an attractive potential that has a finite well at short distances. This means that classically bound states can form. In particular, Eqn.~(\ref{eq:KelbgPotentialDimensionless}) and Fig.~\ref{Kelbg_varyTheta} show that higher values of $\Gamma$ and $\Theta$ will lead to a deeper potential well. This means that bound states will be more likely to form and have a higher impact on the MD simulations with higher $\Gamma$ and $\Theta$ values. The impact of these bound states on the calculation of the bremsstrahlung emission coefficient will be discussed in more detail in Sec.~\ref{subsec:ImpactofBoundStates}. Here we note that the timestep required to accurately resolve bound states limits the values of $\Gamma$ and $\Theta$ that are achievable in the MD simulations. The plasma conditions shown in Table~\ref{tab:table2} were chosen for computational feasibility and to span a wide range of temperatures and densities relevant to hydrogen plasmas.

\begin{table}[]
\caption{\label{tab:table2} Plasma temperatures and total number densities $n = n_{i} + n_{e}$ corresponding to values of $\Gamma$ and $\Theta$ for $Z = 1$.}
\begin{ruledtabular}
\begin{tabular}{cccc}
 $\Gamma$ &$\Theta$ &Temperature [$eV$]
 &Density [$m^{-3}$]\\
\hline
0.01& 468 & 500 & $10^{28}$\\
0.01& 46.8 & 5000 & $10^{31}$\\
0.01& 4.68 & 50000 & $10^{34}$\\
0.1& 46.8 & 50 & $10^{28}$\\
0.1& 4.68 & 500 & $10^{31}$\\
0.1& 0.468 & 5000 & $10^{34}$\\
1& 4.68 & 5 & $10^{28}$\\
1& 0.468 & 50 & $10^{31}$\\
\end{tabular}
\end{ruledtabular}
\end{table}

The bremsstrahlung emission coefficient was calculated from the simulations using a well known formula for the non-relativistic dipole emission from a collection of charges~\cite{befeki1966radiation}
\begin{eqnarray}
    \label{eq:emissioncoefficientMD}
    j(\omega) = \frac{1}{12\pi^2\epsilon_{0}c^3V\mathcal{T}}\left|\int_{-\infty}^{\infty}\sum_{j}\frac{q_{j}}{m_{j}}\mathbf{F}_{j}(t)e^{-i\omega t}dt\right|^{2} ,
\label{collection}
\end{eqnarray}
where $\mathbf{F}_{j}(t)$ is the force on each particle $j$ at time $t$, $V$ is the simulation volume, and $\mathcal{T}$ is the total runtime. The emission coefficient $j(\omega)$ represents the power emitted per unit frequency per unit solid angle and is given in units of W/m$^{3}$/Hz/sterr. Because the bremsstrahlung power spectrum $W(\omega)=V\mathcal{T}j(\omega)$ and the emission coefficient are only related by the simulation volume and total runtime, this paper will refer to $j(\omega)$ interchangeably as the spectrum and emission coefficient.

\subsection{\label{subsec:SimulationResults} Simulation Results}

\begin{figure}\label{SimResults}
    \includegraphics[width=0.5\textwidth]{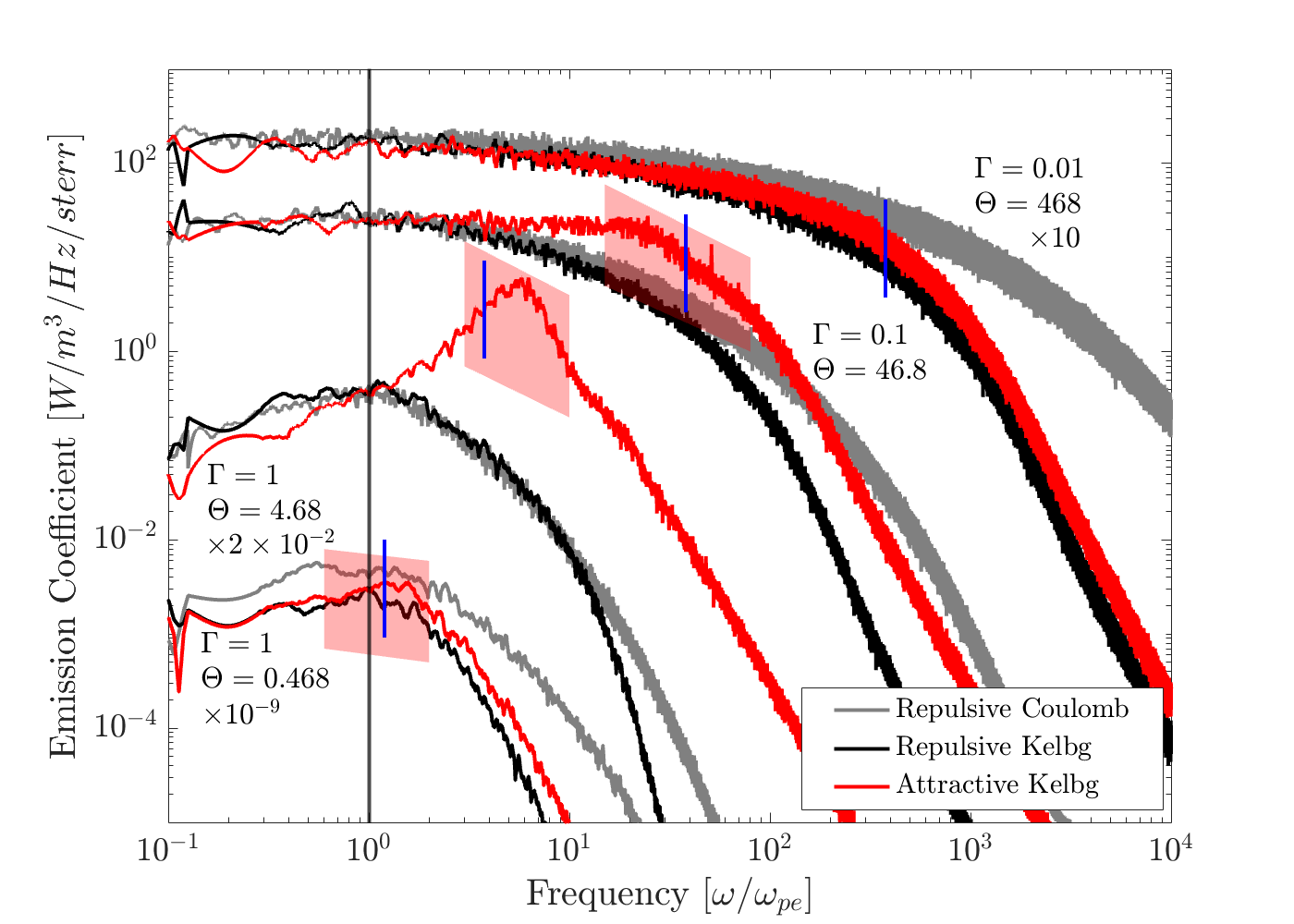}
    \caption{\label{SimResults} Emission coefficient for a range of coupling strengths and degeneracy parameters obtained from MD simulations using Eq.~(\ref{eq:emissioncoefficientMD}). The solid vertical line is at the electron plasma frequency ($\omega_{pe}$). At each value of $\Gamma$ and $\Theta$ there are three curves corresponding to three different simulation cases: a positron-ion plasma interacting through the repulsive Coulomb potential (grey), a positron-ion plasma interacting through the repulsive Kelbg potential (black), and an electron-ion plasma interacting through the attractive Kelbg potential. The electron-electron (positron-positron) and ion-ion interactions were still modeled through the Coulomb potential in all cases. The MD data for $\Gamma = 0.01, 1, 1$ and $\Theta = 468,4.68,0.468$ are multiplied by $10,0.02,10^{-9}$ for clarity. The solid blue vertical lines represent the decay frequencies given by Eq.~(\ref{eq:omega_max_kelbg}). The red shaded regions show the frequency range where bound states contribute to the emission spectrum (see Sec.~\ref{subsec:ImpactofBoundStates}).}
    \end{figure}
Spectra calculated from various MD simulations are shown in Fig.~\ref{SimResults}. In order to reduce high frequency noise and isolate the general trends, a third-degree Savitzky-Golay filter was used in post-processing the data.~\cite{savitzky1964smoothing}. Due to the large ion-electron mass ratio $m_{i}/m_{e}\approx1836$, the sum in Eq.~(\ref{eq:emissioncoefficientMD}) is dominated by the forces on electrons. Thus, the qualitative behavior of the spectra is best understood by considering the electron and positron collisional timescales.
%present in the plasma. 
A physical understanding is developed through examining three cases: repulsive Coulomb interactions, repulsive Kelbg interactions, and attractive Kelbg interactions.  

\subsubsection{High Frequency Decay}
In the high frequency limit ($\omega\gg\omega_{pe}$) the emission spectra decay rapidly toward zero. This decay is related to the timescale of individual binary collisions between an electron (positron) and an ion. Physically, this close interaction timescale is $\tau_{\min}\approx b_{\min}/v_{Te}$, where $v_{Te} = \sqrt{2k_\textrm{B}T/m_{e}}$ is the electron thermal speed and $b_{\min}$ is the distance of closest approach. For the repulsive Coulomb system, the distance of closest approach is the Landau length $b_{\min} = r_{L} = Zq^{2}/(4\pi\epsilon_{0}k_\textrm{B}T)$, which is found by setting $|U_{C}(r_\textrm{L})| = k_{B}T$. This gives a decay frequency $\omega_\textrm{max,RC}\approx\tau_\textrm{min,RC}^{-1} \approx v_{Te}/r_\textrm{L}$ or
\begin{eqnarray}
\label{eq:omega_max}
    \omega_\textrm{max,RC}/\omega_{pe} \approx 2/(\sqrt{3}\Gamma^{3/2}) .
\end{eqnarray}
for $Z = 1$. Physically, the repulsive Coulomb spectrum decays because there is a limit to how close a positron and ion can approach each other.

%Although this gives an estimate of where the repulsive Coulombic spectra decay, it is not an absolute cutoff. This is because Eq.~(\ref{eq:omega_max}) assumes all electrons move at the thermal speed. Some electrons will move slower or faster, leading to a gradual decay of the spectrum. 
The distance of closest approach is modified for the repulsive Kelbg potential and is found by setting $|U_\textrm{K}(r_\textrm{L,K})| = k_{B}T$. This potential also introduces the Kelbg scale length $\lambda$, which puts an additional upper bound on the force an electron can feel. If $r_\textrm{L,K}$ exists and $r_\textrm{L,K} > \lambda$, then there will be a decay at a frequency $\omega_\textrm{max,RK}\approx\tau_{min,RK}^{-1} \approx v_{Te}/r_\textrm{L,K}$. This expectation is corroborated by the repulsive Kelbg spectrum for $\Gamma = 1$ and $\Theta = 4.68$ in Fig.~\ref{SimResults}. Here $r_\textrm{L,K}/\lambda \approx 3.30$ and there is a decay of the repulsive Kelbg spectrum at $\omega_\textrm{max,RK}\approx1.15$. 

There will also always be a decay at a frequency associated with the Kelbg scale length. This is given by $\omega_\textrm{max,K}\approx\tau_\textrm{min,K}^{-1} \approx v_{Te}/\lambda$ or
\begin{eqnarray}
\label{eq:omega_max_kelbg}
    \omega_\textrm{max,K}/\omega_{pe} \approx 1.759\sqrt{\frac{\Theta}{\Gamma}}.
\end{eqnarray}
For the following $\Gamma= \left(0.01, 0.1, 1, 1\right)$ and $\Theta =\left(468, 46.8, 4.68, 0.468\right)$ combinations, Eq.~(\ref{eq:omega_max_kelbg}) predicts a decay at frequencies $\omega_\textrm{max,K}/\omega_{pe} \approx 380, 38, 3.8,$ and $1.2$. These values qualitatively agree with the MD calculations in Fig.~\ref{SimResults}.

For the attractive Kelbg potential, there will only be a sharp decay around a maximum frequency given by Eq.~(\ref{eq:omega_max_kelbg}). In the case of the attractive spectra there are two main differences from the repulsive case: the electron dynamics during a collision with an ion and the formation of classically bound electron-ion states. Firstly, the sign of the potential will cause the electron dynamics and specifically the time history of the force to be slightly different between the attractive and repulsive collisions. This is shown for the case of $\Gamma = 0.01$ and $\Theta = 468$ in Fig.~\ref{SimResults}, where the black and red curves show a slight deviation from each other at high frequency. Electrons under an attractive force will, on average, be able to approach closer to an ion than positrons will. This means the emission from attractive interactions will always be higher than that in the repulsive case.  Secondly, the formation of classically bound states is unique to attractively interacting systems and leads to distinct peaks in the attractive simulations. This feature will be discussed in more detail in Sec.~\ref{subsec:ImpactofBoundStates}.

\subsubsection{Low Frequency Decay}
In the low frequency limit ($\omega\ll\omega_{pe}$) the emission spectra also decay to zero. This is associated with the transition to a fluid dynamical timescale where the net acceleration of the plasma is zero. The characteristic decay at low frequencies is associated with the Coulomb collision timescale $\tau_{ei} = \nu_{ei}^{-1}$ where $\nu_{ei}$ is the positron-ion or electron-ion collision frequency. In the case of the attractive Kelbg simulations, a calculation of $\nu_{ei}$ differs from the repulsive Coulomb case through both the sign of the interaction, known as the Barkas effect,~\cite{shaffer2019thebarkas} and the non-Coulombic short range part of the potential. For conditions of $\Gamma \approx 1$ and $Z = 1$ the Barkas effect is expected to increase $\nu_{ei}$ by around $100\%$. However, this increase in $\nu_{ei}$ is countered by the decrease due to the soft core. As $\Theta$ decreases, the plateau in the Kelbg potential extends further out in radial distance and would be expected to lower the value of $\nu_{ei}$.\cite{shaffer2019thebarkas} For a fixed $\Theta$, the collision frequency approaches $\omega_{pe}$ with increasing $\Gamma$, so the value of the low frequency cutoff increases with $\Gamma$. 

To illustrate this point, a rough estimate of the decay frequency is given by the Coulomb collision timescale $\omega_{\min} = \tau_{ei}^{-1}$ so that in the weakly coupled limit
\begin{eqnarray}
    \label{eq:omega_min}
    \omega_{\min}/\omega_{pe} \approx \Gamma^{3/2} \ln \Lambda / \sqrt{3\pi}, 
\end{eqnarray}
where $\ln\Lambda$ is the Coulomb logarithm. For $\Gamma =0.01$ and $0.1$, this gives $\omega_{\min}/\omega_{pe} \approx 2\times 10^{-3}$ and $3\times10^{-2}$, respectively. It is not possible to compare these predictions with MD because the simulations only resolve frequencies greater than approximately $10^{-2}\omega_{pe}$. Using a fit to MD data from Ref.~\onlinecite{daligault2012diffusion}, the Coulomb logarithm can be approximated as $\ln \Lambda \approx 0.65\ln(1+2.15/(\sqrt{3}\Gamma^{3/2}))$ for $\Gamma \lesssim 20$. For $\Gamma = 1$, this gives a decay frequency of $\omega_{\min}/\omega_{pe} \approx 0.17$. Figure~\ref{SimResults} supports the point that the spectra for $\Gamma = 1$ decay closer to $\omega_{pe}$ than the spectra for the $\Gamma  = 0.01$ and $0.1$ cases. The quantitative details of the low frequency features are discussed at more length in Sec.~\ref{subsec:LF}.

\subsubsection{\label{subsec:ImpactofBoundStates}Impact of Bound States}

    \begin{figure}[h!]\label{BoundStates}
    \includegraphics[width=0.5\textwidth]{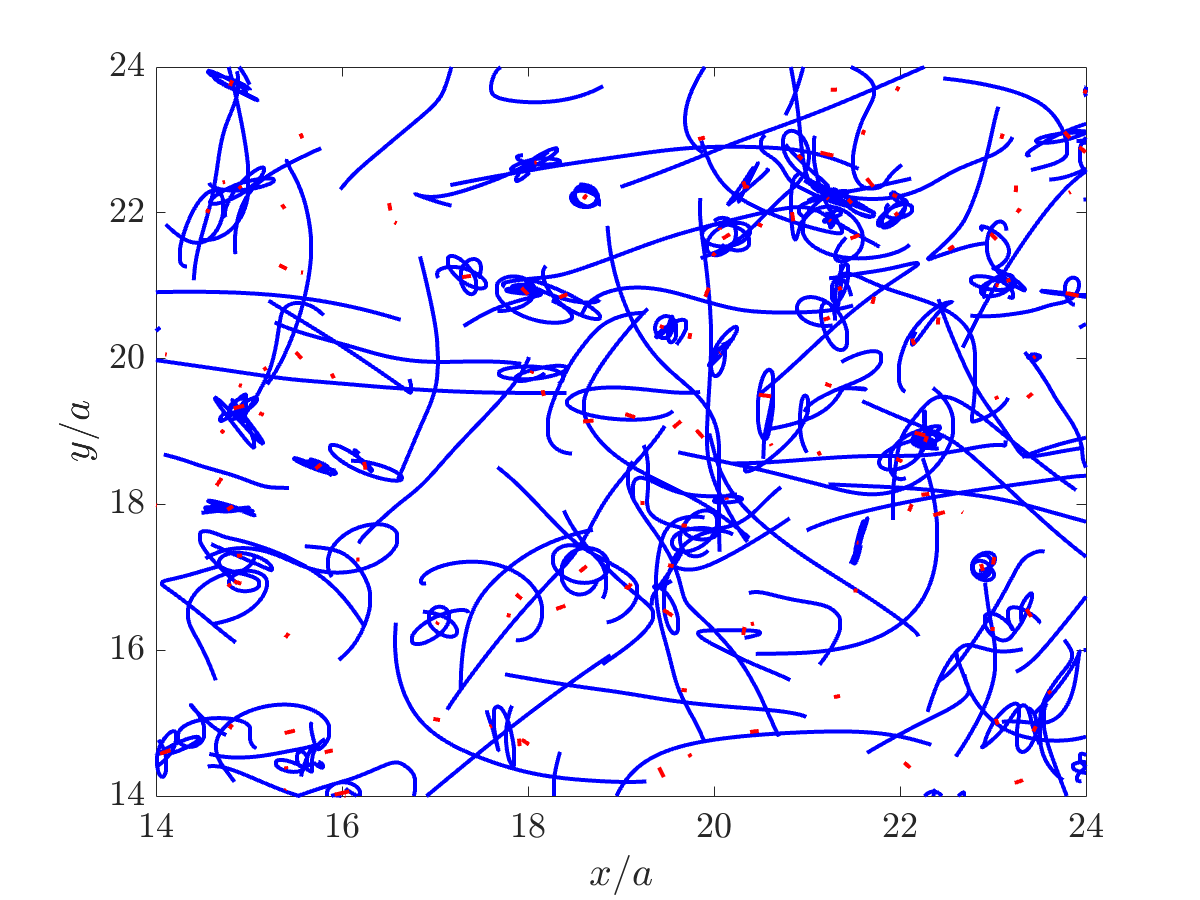}
    \caption{\label{BoundStates} Electron (blue) and ion (red) trajectories over a time period of $3\omega_{pe}^{-1}$ for an MD simulation with $\Gamma = 1$ and $\Theta = 4.68$. The three dimensional trajectories were projected into two dimensions for visualization purposes.}
    \end{figure}
    
\begin{figure}[h!]\label{RDFsBound}
    \includegraphics[width=0.5\textwidth]{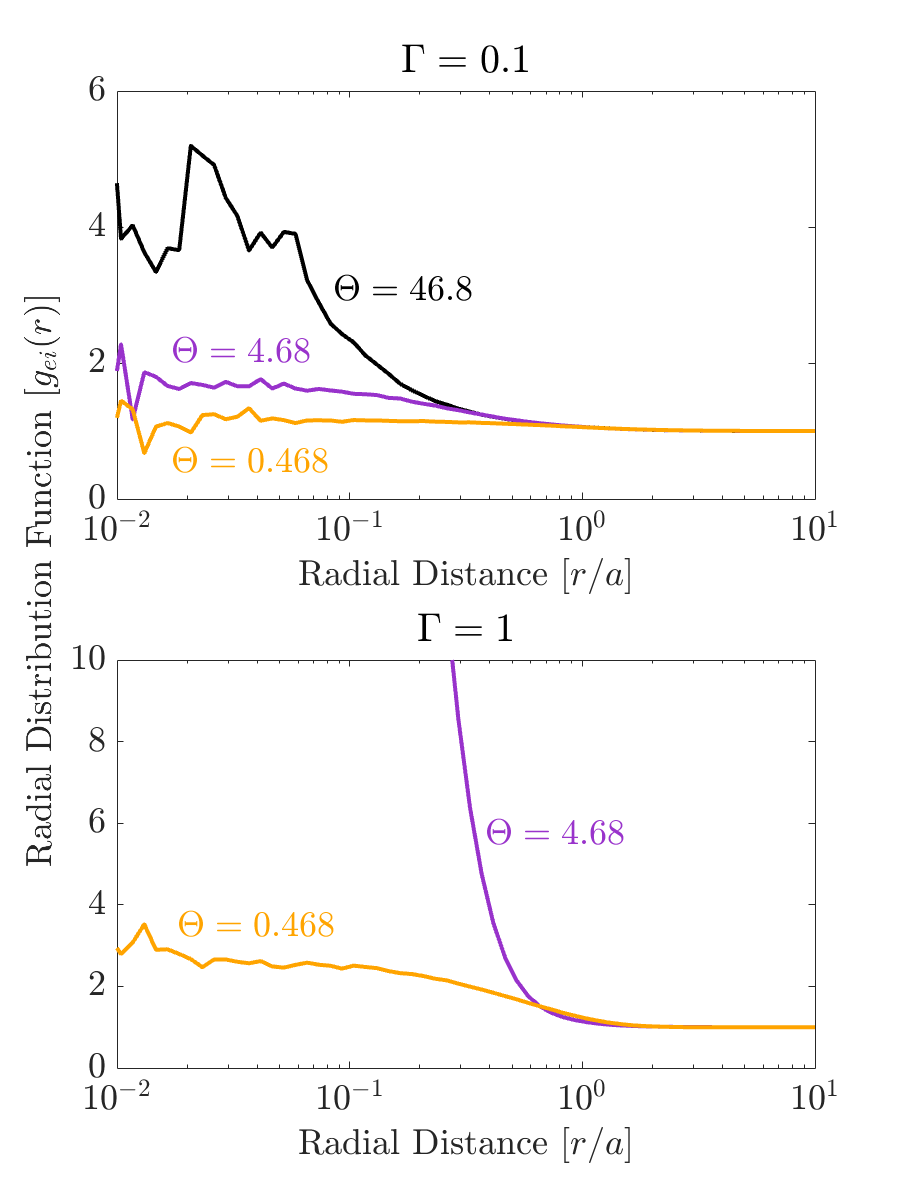}
    \caption{\label{RDFsBound} Electron-ion radial distribution functions calculated from MD simulations using Eq.~(\ref{eq:grMD}).}
    \end{figure}
    
A unique aspect of attractively interacting electrons and ions is that classical bound states form from three-body recombination.\cite{tiwari2017thermodynamic} A large fraction of bound states is observed in the MD simulations when $\Gamma$ and $\Theta$ are sufficiently large, corresponding to conditions where the Kelbg potential reaches larger negative values at close distances (see Fig.~\ref{Kelbg_varyTheta}).

The existence of bound states can be deduced by directly looking at the particle trajectories (see Fig.~\ref{BoundStates}) or by examining the electron-ion radial distribution functions (see Fig.~\ref{RDFsBound}). The quantity $4\pi r^{2}ng_{ei}(r)dr$ gives the average number of electron-ion pairs with a radial separation in the range $r = |{\mathbf r}_{i} - {\mathbf r}_{e}|$ to $r + dr$. The electron-ion radial distribution function was calculated from the MD simulations using the following formula\cite{mithen2012a}
\begin{eqnarray}
    \label{eq:grMD}
    g_{ei}(r_\textrm{eff}) = \frac{1}{N_\textrm{steps}N_{e}n_{e}}\sum_{i=1}^{N_\textrm{steps}}\frac{N_\textrm{pairs}(r,\Delta r)}{V(r,\Delta r)},
\end{eqnarray}
where $N_\textrm{steps}$ is the number of time steps, $N_{e}$ is the number of electrons, and $n_{e}$ is the number density of electrons. The quantity $N_\textrm{pairs}(r,\Delta r)$ represents the number of electron-ion pairs with a radial separation between $r$ and $r+\Delta r$ and $V(r,\Delta r)$ is the volume of the enclosed spherical shell. The effective radius is defined as\cite{mithen2012a,poll1988one} 
\begin{eqnarray}
    \label{eq:effectiveradius}
    r_\textrm{eff} = \frac{3}{4}r\left[\frac{\left(1+\Delta r/r\right)^{4}-1}{\left(1+\Delta r/r\right)^{3}-1}\right].
\end{eqnarray}

The results shown in Fig.~\ref{RDFsBound} used $N_\textrm{steps} = 10^{4}$ with particle position data generated at time steps of $0.1\omega_{pe}^{-1}$. The calculations used logarithmically spaced radial points to resolve the small $r$ behavior. The noise at very small values of $r$ has to do with the finite value of $N_\textrm{steps}$ used to generate the data. Regardless, it is clear that the value of $g_{ei}(r)$ reflects the soft-core nature of the electron-ion Kelbg potential at short distances. More importantly, the value of $g_{ei}(r)$ at short distances is much larger for higher $\Gamma,\Theta$ pairs. High values of $g_{ei}(r)$ at close distances indicates an increased probability of finding an electron-ion pair. 

Classically bound electron-ion states show up as peaks in the emission coefficient.  Figure~\ref{SimResults} shows three bound state peaks. Two obvious peaks occur for the case of $\Gamma = 0.1$ and $\Theta = 46.8$ around $\omega/\omega_{pe}\approx23$ and and for the case of $\Gamma = 1$ and $\Theta = 4.68$ around $\omega/\omega_{pe}\approx6$. Another smaller peak occurs for $\Gamma = 1$ and $\Theta = 0.468$ around $\omega/\omega_{pe}\approx1.5$  In order to predict the location of these peaks, one can make the approximation that all bound electrons are caught in a perfectly circular orbit around an ion. Then, the force balance for each orbiting electron $i$ is
\begin{eqnarray}
    \label{eq:forcebalanceBound}
    \frac{dU_\textrm{K}}{dr}\Big|_{r=r_{i}} = \frac{m_{e}v_{i}^{2}}{r_{i}}, 
\end{eqnarray}
where the left hand side is the electron-ion force due to the Kelbg potential, the right hand side is the centripetal force, and $v_{i}$ and $r_{i}$ represent the orbit velocity and radius respectively. A circular orbit will also correspond to a specific angular frequency $\omega_{i} = v_{i}/r_{i}$ at which emission will occur. 

\begin{figure}[h!]\label{Prediction}
    \includegraphics[width=0.5\textwidth]{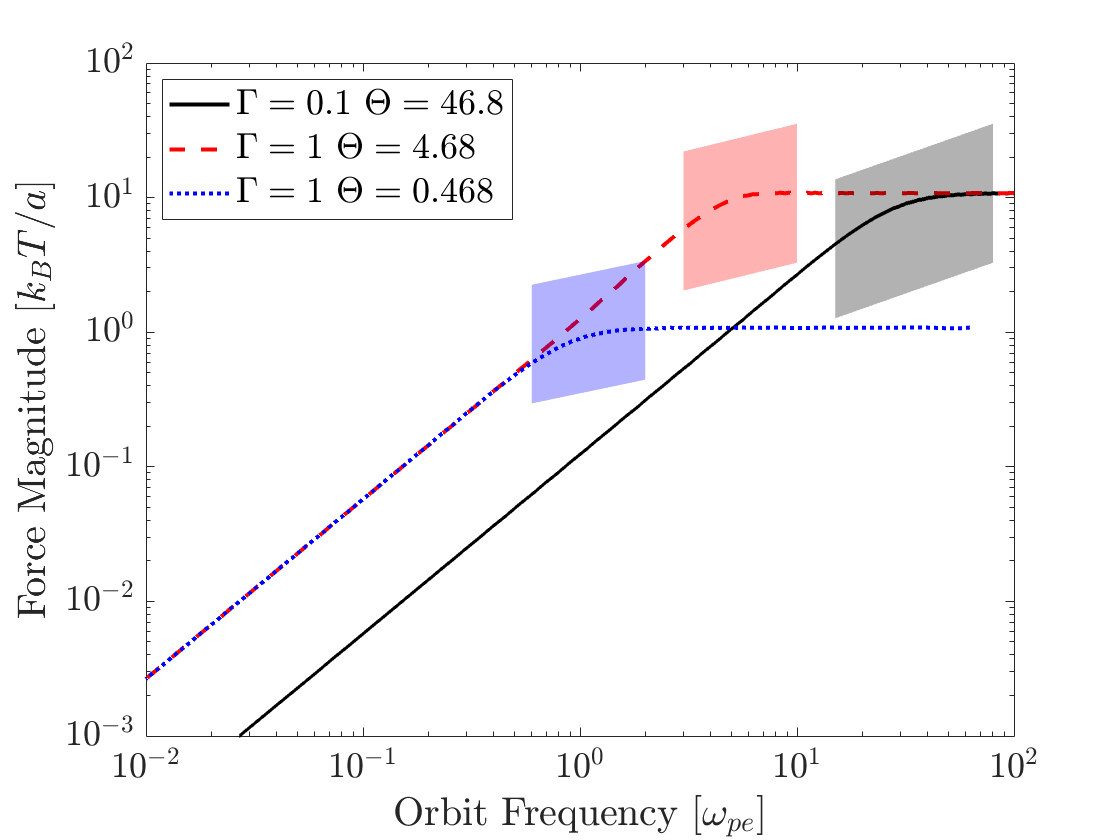}
    \caption{\label{Prediction} The force magnitude and orbit frequencies for an electron in a circular orbit around an ion calculated from Eq.~(\ref{eq:forcebalanceBound}). The shaded regions denote the approximate frequency boundaries that are important for bound state emission. These regions are bounded on the left by a sharp decrease in the force and on the right by a decreasing probability of bound electrons.}
    \end{figure}

Solving Eq.~(\ref{eq:forcebalanceBound}) for the possible values of $v_{i},r_{i},$ and $\omega_{i}$ gives the relative importance of different orbit frequencies. Electrons with smaller orbit radii $r_{i}$ will have a larger orbit frequency $\omega_{i}$, and vise versa. Figure~\ref{Prediction} shows the force on each electron as a function of orbit frequency. The bremsstrahlung spectrum of each bound electron in a circular orbit is simply related to the square of this force. For low orbit frequencies, the force drops off rapidly and becomes negligible. For high orbit frequencies, the force becomes constant per the shape of the Kelbg potential. For high orbit frequencies, bound electrons are also increasingly close to their respective ion. Figure~\ref{RDFsBound} shows that $g_{ei}(r)$ plateaus at small values of $r$. The average number of particles in a spherical shell ($4\pi r^{2}ng_{ei}(r)dr$) within this plateau drops rapidly as $r\xrightarrow{}0$. Thus, the orbit frequencies where the force becomes constant are increasingly less probable to be occupied by bound electrons. This physical argument restricts the important frequencies for bound electron emission to a small region between the sharp decrease and the plateau (see Fig.~\ref{Prediction}). For $\Gamma = 0.1$ and $\Theta = 46.8$ this corresponds to orbit (emission) frequencies in the range $\omega_{i}/\omega_{pe}\approx 15-80$. For $\Gamma = 1$ and $\Theta = 4.68$ this corresponds to emission frequencies in the range $\omega_{i}/\omega_{pe}\approx 3-10$. For $\Gamma = 1$ and $\Theta = 0.468$ this corresponds to emission frequencies in the range $\omega_{i}/\omega_{pe}\approx 0.6-2$. This agrees well with the location of the bound state peaks in Fig.~\ref{SimResults}.

However, such classically bound electrons are not likely to contribute to the observed bremsstrahlung emission of a physical system at the $\Gamma$ and $\Theta$ conditions considered here. This is because the positions of the electrons in the bound states correspond to the lowest energy levels of the hydrogen atom. The expectation values of the radial coordinate of electron wavefunctions in hydrogen is~\cite{sakurai2011modern}
\begin{eqnarray}
    \label{eq:ExpectationValue}
    \langle\psi_\textrm{n,l}|\hat{r}/a|\psi_\textrm{n,l}\rangle = \left(\frac{8}{9\pi}\right)^{2/3}\frac{\left[3n^{2}-l(l+1)\right]}{\Gamma\Theta}.
\end{eqnarray}
For $\Gamma = 0.1$ and $\Theta = 46.8$, this expression gives $\langle\psi_\textrm{1,0}|\hat{r}/a|\psi_\textrm{1,0}\rangle\approx0.28$. The region for bound state emission in Fig.~\ref{Prediction} corresponds to orbit radii in the range $r_{i}/a\sim0.03-0.13$. Higher combinations of $n,l$ give larger expectation values, so it is reasonable to conclude that the radial bounds in Fig.~\ref{Prediction} correspond best to the hydrogen ground state. This is true of the other $\Gamma,\Theta$ pairs and suggests that these deeply bound states will contribute to line emission but would not be part of the bremsstrahlung spectrum. 

\begin{figure}[h!]\label{facf}
    \includegraphics[width=0.5\textwidth]{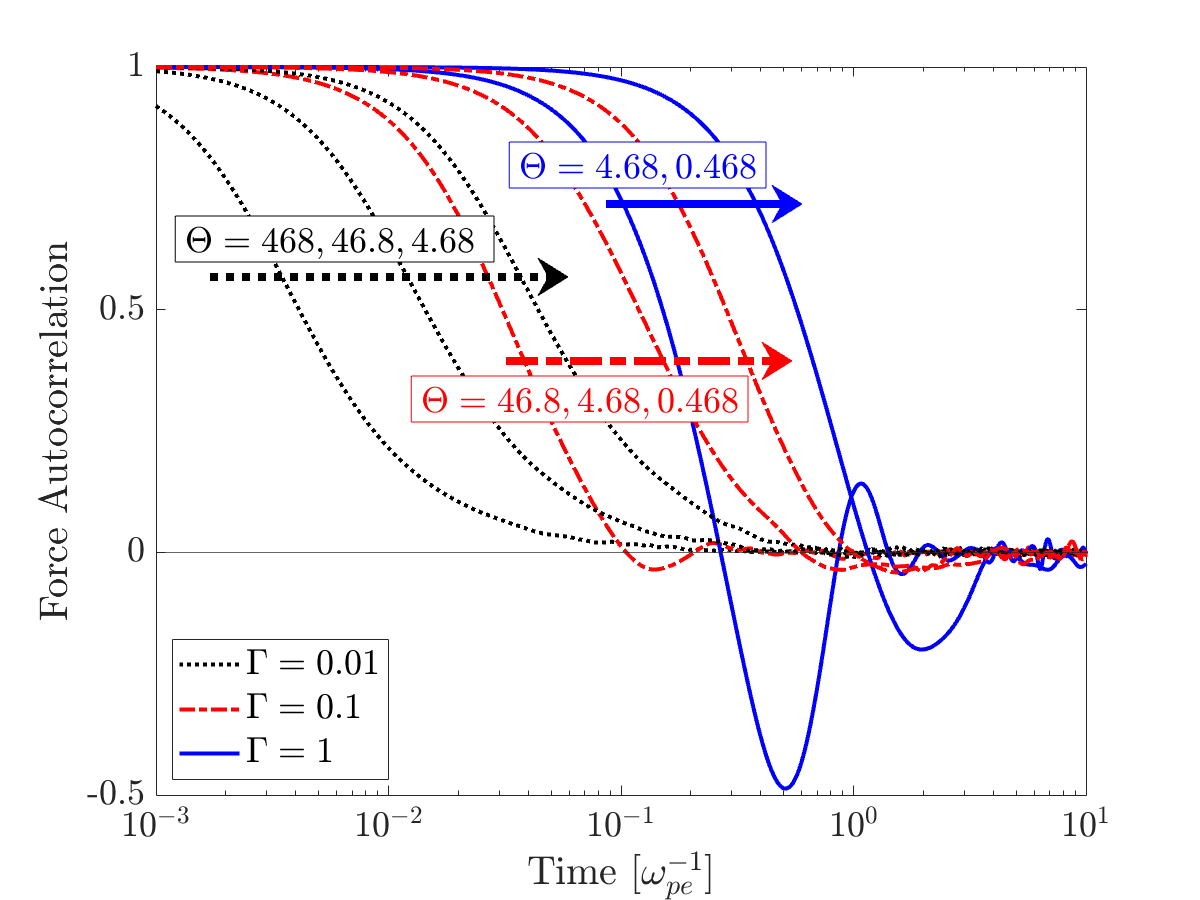}
    \caption{\label{facf} Force autocorrelation functions calculated from MD simulations using Eq.~(\ref{eq:emissioncoefficientFACFtimeave}). The curves are normalized to their value at $t = 0$.}
\end{figure}

\begin{figure*}[t!]\label{facfvsMD}
\includegraphics[width=1.0\textwidth]{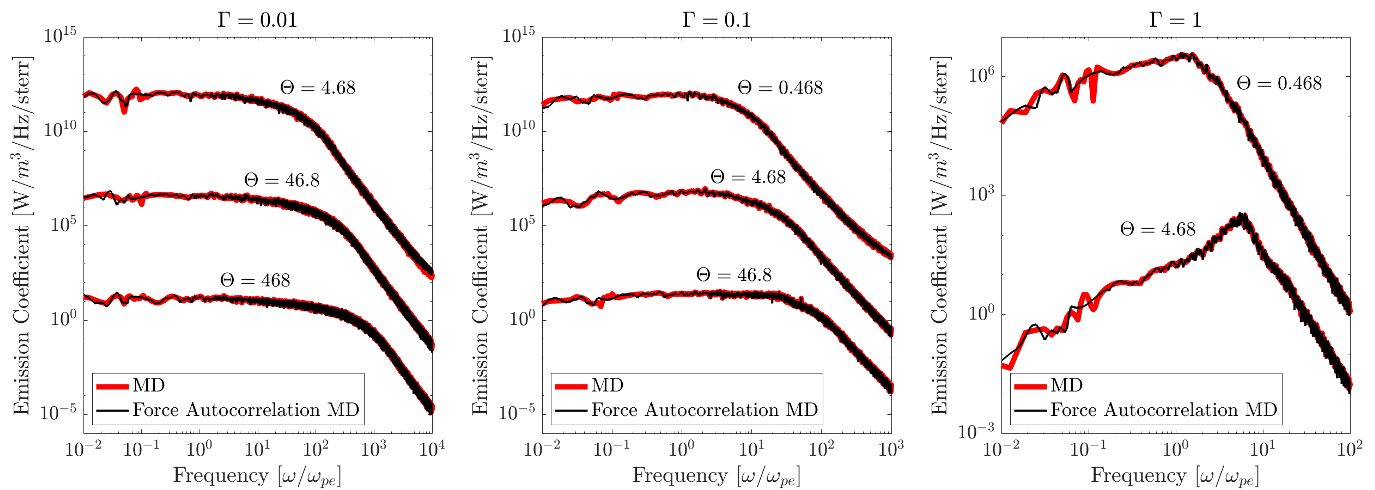}
    \caption{\label{facfvsMD} Emission coefficient calculated from MD simulations using Eq.~(\ref{eq:emissioncoefficientMD}) compared to calculations using Eq.~(\ref{eq:emissioncoefficientFACF}). Good agreement across all frequencies supports using the autocorrelation formalism to describe bremsstrahlung emission.}
\end{figure*}

\subsection{\label{subsec:RelationtoAutocorrelationFunctions}Relation to the Force Autocorrelation Function}

The bremsstrahlung emission coefficient defined by Eq.~(\ref{eq:emissioncoefficientMD}) can alternatively be written as the Fourier transform of a force autocorrelation function. This relation shows the connection between the bremsstrahlung emission coefficient and the dynamic electrical conductivity, as derived from linear response theory (see Chapter 7.6 of Ref.~\onlinecite{hansen2013theory}), and serves as the basis for modeling different frequency regimes in mean force emission theory. 

Applying the Wiener-Khinchin theorem to Eq.~(\ref{eq:emissioncoefficientMD}), the emission coefficient can be written
\begin{eqnarray}
\label{eq:emissioncoefficientFACF}
    j(\omega) = \frac{1}{6\pi^2\epsilon_{0}c^{3}V}\mathrm{Re}\biggl\{\int_{0}^{\infty}\langle\dot{\mathbf{J}}(t)\cdot\dot{\mathbf{J}}(0)\rangle e^{-i\omega t}dt\biggr\}
\end{eqnarray}
where $\langle...\rangle$ represents an ensemble average and $\dot{\mathbf{J}} = \sum_{j}(q_{j}/m_{j})\mathbf{F}_{j}$ is the sum of the total forces in the system, which is the time derivative of the electrical current density $\mathbf{J} = \sum_j q_j \mathbf{v}_j$. Applying the property of time correlations that $-d^2 \langle A(t)B^*\rangle/dt^2 = \langle \dot{A}(t) \dot{B}^*\rangle$, it is seen that the emission coefficient is related to the dynamic conductivity ($\sigma(\omega)$) as 
\begin{eqnarray}
\label{eq:jsigmaconnection}
    j(\omega) = \frac{\omega^{2}k_{B}T}{2\pi^{2}c^{3}\epsilon_{0}}\mathrm{Re}\bigl\{\sigma(\omega)\bigr\},
\end{eqnarray}
where
\begin{eqnarray}
\label{eq:dynamicconductivityFACF2}
    \nonumber\mathrm{Re}\bigl\{\sigma(\omega)\bigr\} = \frac{1}{3k_{B}TV}\mathrm{Re}\biggl\{\int_{0}^{\infty}\langle\mathbf{J}(t)\cdot\mathbf{J}(0)\rangle e^{-i\omega t}dt\biggr\}\\
\end{eqnarray}
is the Green-Kubo relation for the dynamical electrical conductivity coefficient. 

In quantum linear response theory,~\cite{statistical1957kubo,dufty2018kg} the emission coefficient is often expressed as
\begin{equation}
    j(\omega) = B(\omega) \alpha(\omega) 
    \end{equation}
where 
\begin{eqnarray}
\label{eq:blackbody}
    B(\omega) = \frac{\hbar\omega^{3}}{2\pi^{2}c^{2}}\frac{1}{e^{\hbar\omega/k_{B}T}-1}
\end{eqnarray}
is the Planck blackbody spectrum and 
\begin{eqnarray}
\label{eq:absorptioncoeff}
    \alpha(\omega) = \frac{\mathrm{Re}\bigl\{\sigma(\omega)\bigr\}}{c\epsilon_{0}},
\end{eqnarray}
is the absorption coefficient. 
The real part of the dynamic conductivity is given by the Kubo relation
\begin{eqnarray}
\label{eq:dynamicconductivityFACF1}
    \nonumber\mathrm{Re}\bigl\{\sigma(\omega)\bigr\} = \frac{1-e^{-\frac{\hbar\omega}{k_{B}T}}}{3\hbar\omega V}\mathrm{Re}\biggl\{\int_{0}^{\infty}\langle\Hat{\mathbf{J}}(t)\cdot\Hat{\mathbf{J}}(0)\rangle e^{-i\omega t}dt\biggr\},\\
\end{eqnarray}
where $\Hat{\mathbf{J}} = \sum_{j}q_{j}\Hat{\mathbf{v}}_{j}$ is the current operator.
%where $\Hat{\mathbf{J}} = \sum_{j}\frac{q_{j}}{2m_{j}}\left\{\Hat{\mathbf{p}}_{j},\delta\left(\Hat{\mathbf{r}}-\Hat{\mathbf{r}}_{j}\right)\right\}$ is the current operator and $\left\{...\right\}$ represent Poisson brackets.
Thus, Eq.~(\ref{eq:jsigmaconnection}) can simply be obtained from the classical limit ($\hbar\omega\ll k_{B}T$) of the common quantum formula. 
In the classical limit the operators ($\hat{\mathbf{J}}$) become classical variables ($\mathbf{J}$) and $\hbar\omega\ll k_{B}T$ so that Eq.~(\ref{eq:dynamicconductivityFACF1}) equates to Eq.~(\ref{eq:dynamicconductivityFACF2}). 
Similarly, the classical limit of the blackbody spectrum is the Rayleigh-Jeans law: $B_\textrm{c} (\omega) = \omega^2k_\textrm{B} T/(2\pi^2 c^2)$, leading to the prefactor in Eq.~(\ref{eq:jsigmaconnection}). 

In order to fully include medium effects, the definitions of the absorption and emission coefficients would include the real part of the refractive index $n(\omega)$ such that they become $\alpha(\omega)/n(\omega)$ and $n(\omega)j(\omega)$ respectively. The real part of the refractive index of a material can be calculated from the real and imaginary parts of the complex dielectric function $\hat{\epsilon}(\omega) = \epsilon(\omega)/\epsilon_{0}$ such that
\begin{eqnarray}
    \label{eq:refractiveIndex}
    n(\omega) = \sqrt{\frac{\textrm{Re}\left\{\hat{\epsilon}(\omega)\right\}+|\hat{\epsilon}(\omega)|}{2}}.
\end{eqnarray}
The complex dielectric function can be also obtained from the real part of the dynamic conductivity. This is done by applying the equation $\textrm{Im}\left\{\hat{\epsilon}(\omega)\right\} = \textrm{Re}\left\{\sigma(\omega)\right\}/(\epsilon_{0}\omega)$ and the Kramers-Kronig relations. This work omits the refractive index and focuses on the emission coefficient in Eq.~(\ref{eq:emissioncoefficientFACF}), but it can be obtained directly from the emission coefficient in this way.

The force autocorrelation function in Eq.~(\ref{eq:emissioncoefficientFACF}) can be calculated from an MD simulation using the ergodic theorem to interpret the ensemble average as a time average such that
\begin{eqnarray}
\label{eq:emissioncoefficientFACFtimeave}
    \langle\dot{\mathbf{J}}(t)\cdot\dot{\mathbf{J}}(0)\rangle = \lim_{T\xrightarrow{}\infty}\frac{1}{T}\int_{0}^{T}\dot{\mathbf{J}}(t^{'})\cdot\dot{\mathbf{J}}(t^{'}+t)dt^{'}.
\end{eqnarray}
The MD data used to perform this calculation was identical to that used to calculate the curves in Fig.~\ref{SimResults} and described in Sec.~\ref{subsec:Setup}. Figure~\ref{facf} shows that the force autocorrelations change qualitatively with coupling strength and degeneracy parameter. As $\Gamma \rightarrow 1$, the force autocorrelations show oscillatory behavior on the order of $\omega_{pe}^{-1}$ that is indicative of strongly correlated motion. Note that the oscillatory behavior for $\Gamma = 0.1,1$ and $\Theta = 46.8,4.68$ are dominated by the formation of classically bound electron-ion states. Figure~\ref{facfvsMD} shows that the expression for the emission coefficient using Eq.~(\ref{eq:emissioncoefficientFACF}) agrees with the calculation from Eq.~(\ref{eq:emissioncoefficientMD}).

\section{\label{sec:theory}Mean Force Emission Theory}

Mean force emission theory~\cite{kinney2024mean} is a generalization of mean force kinetic theory\cite{baalrud2019mean} to a broader range of timescales. In mean force kinetic theory, electron-ion interactions are modeled as binary interactions occurring through a mean force $\Bar{\mathbf{F}}_{ei} = -\nabla w_{ei}\left(r\right)$
where $w_{ei}\left(r\right)$ is the potential of mean force. Physically, the mean force $\Bar{\mathbf{F}}_{ei}$ is the force obtained when an electron and ion are held at a fixed relative position $r = \left|\mathbf{r}_{e}-\mathbf{r}_{i}\right|$ and all other particles are averaged over in a canonical ensemble. The potential of mean force is related to the electron radial distribution function by~\cite{hansen2013theory}
\begin{eqnarray}
    \label{eq:MeanForce}
    w_{ei}\left(r\right) = -k_{B}T\ln{\left[g_{ei}\left(r\right)\right]}.
\end{eqnarray}
The electron-ion radial distribution function can be computed from experimental data, MD simulations,\cite{mithen2012a} or self-consistent numerical iteration of the Ornstein-Zernike equation\cite{hansen2013theory} and hypernetted-chain closure\cite{VANLEEUWEN1959new,baalrud2013effective}. For the following results, we calculate the electron-ion radial distribution function from MD simulations using Eq.~(\ref{eq:grMD}). In order to reduce noise the result is processed using a smoothing spline. It is noted that in the $\Gamma\ll 1$ and $\Theta\gg 1$ limit, the potential of mean force is well approximated by multiplying a Debye-H\"{u}ckel screening term onto the Kelbg potential such that $w_{ei}\left(r\right) \approx U_{K}(r)\exp{\left(-r/\lambda_{D}\right)}$ where $\lambda_{D} = \sqrt{\epsilon_{0}k_{B}T/(nq^{2})}$ is the total Debye length for $Z = 1$.

The theoretical framework for mean force emission theory is based on approximations of Eq.~(\ref{eq:emissioncoefficientFACF}) at timescales corresponding to the high ($\omega\gg\omega_{pe}$) and low ($\omega\ll\omega_{pe}$) frequency limits of the bremsstrahlung spectrum. In the high frequency limit (see Sec.~\ref{subsec:HF}), the spectrum is calculated by solving for the time-history of individual electron trajectories in the presence of the potential of mean force. These short timescale electronic trajectories are summed to obtain the total bremsstrahlung spectrum. In the low frequency limit (see Sec.~\ref{subsec:LF}), a successful model must capture multiple collisions occurring over a longer timescale. To do this, the spectrum is recast in terms of a single electron velocity autocorrelation. Modeling the velocity autocorrelation function as an exponential decay gives the well known `Drude' model of the emission spectrum. A simple comprehensive model bridges these limits to calculate the bremsstrahlung spectrum across all frequencies (see Sec.~\ref{subsec:CM}). Section~\ref{subsec:CM} also argues that the typical high frequency formula with a `Drude' correction at low frequencies cannot capture a peak near the plasma frequency due to strongly correlated motion that occurs on intermediate timescales.

\begin{figure*}[t!]\label{mfetvsMD}
\includegraphics[width=1.0\textwidth]{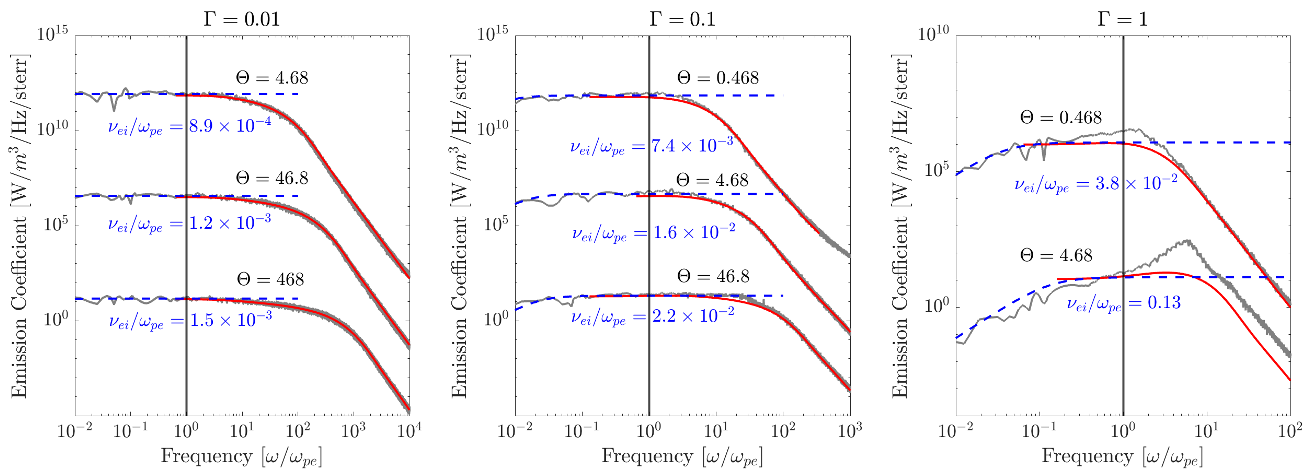}
    \caption{\label{mfetvsMD} High (solid-red) and low (dashed-blue) frequency limits of mean force emission theory compared to MD simulations.}
\end{figure*}

\subsection{\label{subsec:HF}High Frequency Limit}
For a short timescale binary electron-ion interaction, Eq.~(\ref{eq:emissioncoefficientFACF}) can be reduced to finding the time-history of individual electron trajectories during a binary interaction via the potential of mean force. A derivation of this is given in Appendix A of Ref.~\onlinecite{wierling_millat_röpke_2004}. In this high frequency limit, the calculation of the bremsstrahlung emission coefficient is performed by averaging the radiation spectrum from a single electron 
\begin{eqnarray}
    \label{eq:SingleParticleSpectrum}
    \nonumber\mathcal{W}(\omega,v,b) = \frac{q_{e}^{2}}{12\pi^{2}\epsilon_{0}c^{3}m_{e}^{2}}\left|\int_{-\infty}^{\infty}\mathbf{F}_{e}\left[\mathbf{r}(t,v,b)\right]e^{-i\omega t}dt\right|^{2}\\
\end{eqnarray}
over all initial electron speeds ($v$) and impact parameters ($b$) such that the emission coefficient is
\begin{eqnarray}
    \label{eq:HFemissioncoefficient}
    j_{h}(\omega) = 8\pi^{2} n_{i}\int_{0}^{\infty}dv\, v^{3}f_\textrm{M}(v)\int_{0}^{\infty}db\, b\mathcal{W}(\omega,v,b),
\end{eqnarray}
where the cross sectional area of the collision volume is $bdbd\phi = 2\pi bdb$ and the velocity distribution is assumed to be isotropic so that $d^{3}v = 4\pi v^{2}dv$. To compare with the equilibrium MD results the electron velocity distribution $f_\textrm{M}(v)$ is a Maxwellian. The time dependent force on the electron is determined by solving Newton's equation of motion with the potential of mean force. The electron-ion separation ($r$) is taken to be farther away than the screening length of the potential such that $r\gg\lambda_{D}$ at weak coupling and $r\gg a$ for strong coupling. We neglect the ion velocity due to the assumption of thermal equilibrium ($T = T_{e} = T_{i}$) making the thermal velocity of the electrons greater than the ions by a factor of the mass ratio ($\sqrt{m_{i}/m_{e}}$).

Figure~\ref{mfetvsMD} shows that the results of this high-frequency calculation agree well with the molecular dynamics simulations for $\omega > \omega_{pe}$. The two cases that exhibit large amounts of bound electron-ion states, $\Gamma = 0.1, 1$ and $\Theta = 46.8, 4.68$ respectively, show a bound state peak that mean force emission theory cannot capture because it does not model the requisite three-body interactions. Particularly in the case of $\Gamma = 1$ and $\Theta = 4.68$, the bound state peak may be so dominant that it shifts the magnitude of the spectrum calculated from the MD simulations upwards at high frequencies. As in the repulsive case, the high-frequency emission coefficient plateaus to a constant as $\omega\xrightarrow{}0$. This is because the high-frequency limit only treats timescales shorter than the electron-ion binary interaction timescale. In order to capture the low frequency behavior, one needs to include longer timescales associated with multiple electron-ion collisions.

\subsection{\label{subsec:LF}Low Frequency Limit}
The low-frequency behavior is captured by recasting Eq.~(\ref{eq:emissioncoefficientFACF}) in terms of a single electron velocity autocorrelation function. We only briefly outline the steps to do this here, for a more detailed derivation see Ref.~\onlinecite{kinney2024mean}. First, we ignore the forces on ions and cross correlations between different electrons. Then, the emission coefficient can be written as
\begin{eqnarray}
    \label{eq:emissioncoefficientVACF}
    j(\omega) = \frac{n_{e}q^{2}\omega^{2}}{2\pi^{2}\epsilon_{0}c^{3}}\mathrm{Re}\biggl\{\int_{0}^{\infty}Z(t)e^{-i\omega t}dt\biggr\}
\end{eqnarray}
where $Z(t) = (1/3)\langle\mathbf{v}_{e}(t)\cdot\mathbf{v}_{e}(0)\rangle$ is the electron velocity autocorrelation function. Using the Langevin equation, this velocity autocorrelation is modeled as an exponential decay so that $Z(t) = (k_{B}T/m_{e})\exp{(-\nu_{ei}t)}$. Direct electron-electron forces do not contribute to the sum of the forces in the dipole approximation, and thus only the electron-ion collision frequency ($\nu_{ei}$) appears in the result. Then
\begin{eqnarray}
    \label{eq:LFemissioncoefficient}
    j_{l}(\omega) = \frac{\omega_{pe}^{2}k_{B}T}{2\pi^{2}c^{3}}\frac{\nu_{ei}\omega^{2}}{\nu_{ei}^{2}+\omega^{2}}
\end{eqnarray}
which is similar to the Drude form used in quantum calculations of dynamic conductivity and inverse bremsstrahlung.\cite{starrett2016kubo,shaffer2017free,Perrot_1996} The collision frequency is calculated using mean force kinetic theory,~\cite{baalrud2013effective} where the potential of mean force is obtained from the MD simulations. 

Results computed from Eq.~(\ref{eq:LFemissioncoefficient}) agree well with the MD simulation results for $\omega < \omega_{pe}$ (see Fig.~\ref{mfetvsMD}). For fixed $\Gamma$, Fig.~\ref{mfetvsMD} also shows that the value of $\nu_{ei}/\omega_{pe}$ decreases for decreasing $\Theta$. This is because the extended plateau in the Kelbg potential limits the scattering angle of electrons.  At weak coupling, the high and low frequency results are sufficient to describe the spectrum and can be patched together into a comprehensive model.

\subsection{\label{subsec:CM}Comprehensive Model and Plasma Frequency Peak}

\begin{figure*}[t!]\label{DrudevsMD}
\includegraphics[width=1.0\textwidth]{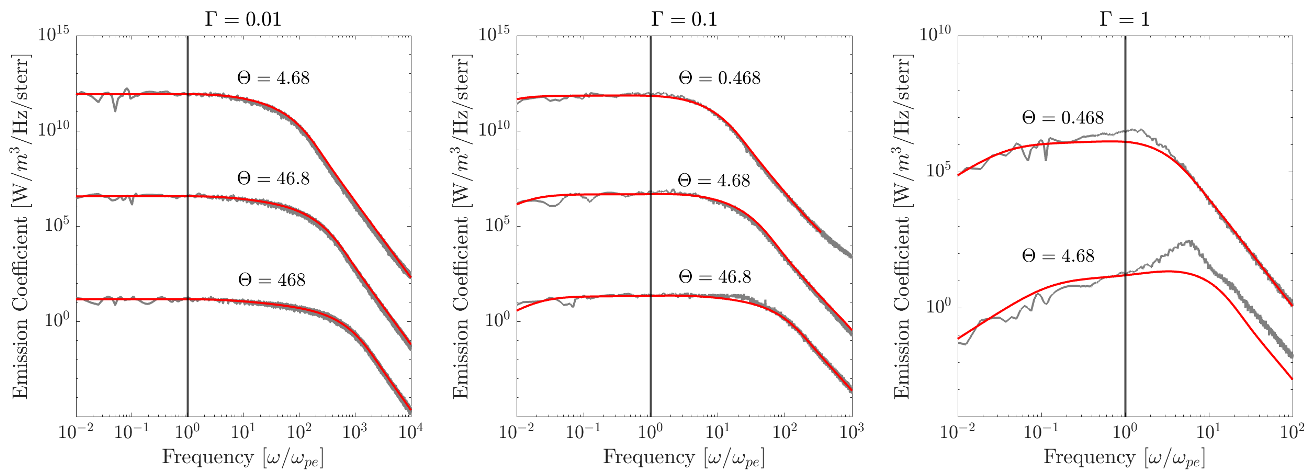}
    \caption{\label{DrudevsMD}Comprehensive solution obtained from Eq.~(\ref{eq:comprehensive}) (red lines) compared to MD simulations (grey lines).}
\end{figure*}

\begin{figure}[t!]\label{RepulsiveandAttractive}
\includegraphics[width=0.5\textwidth]{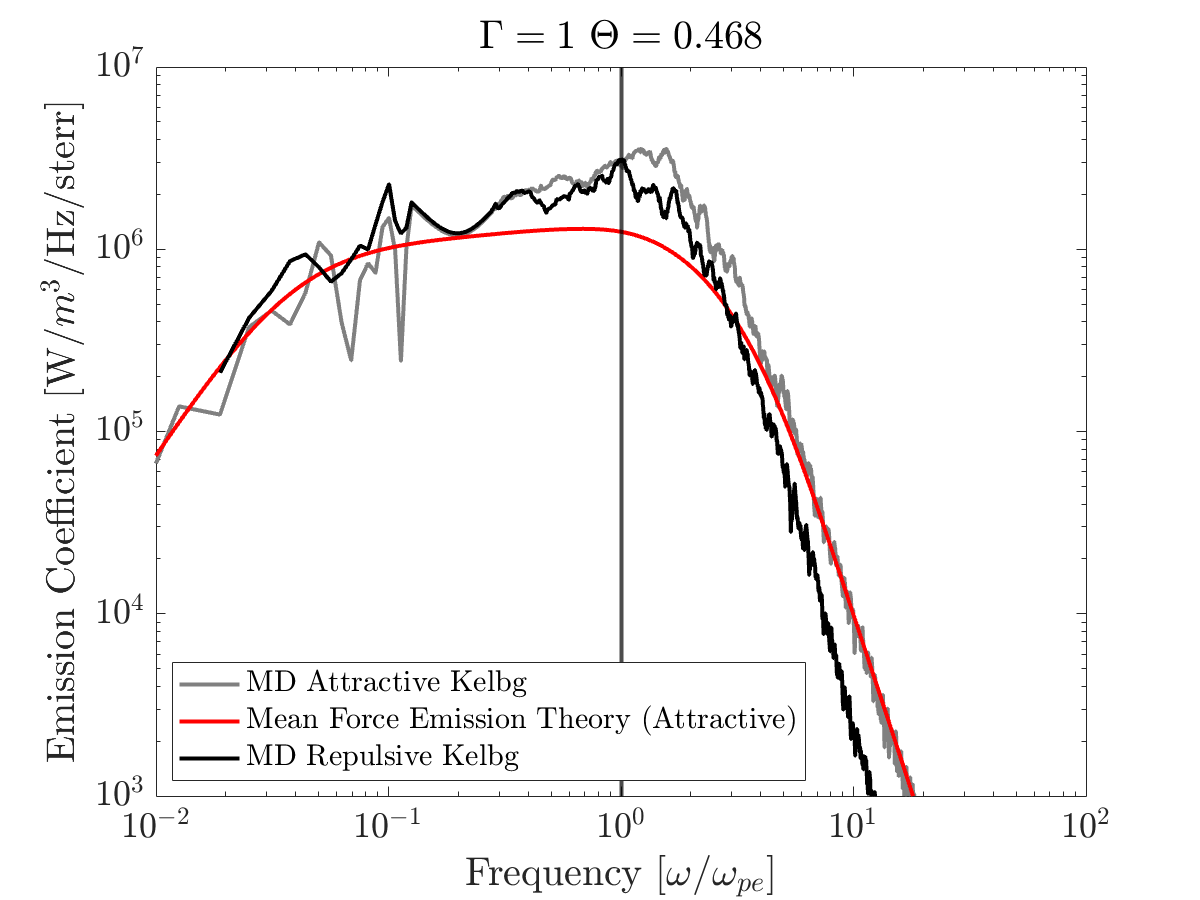}
    \caption{\label{RepulsiveandAttractive} Comprehensive solution obtained from Eq.~(\ref{eq:comprehensive}) compared to MD simulations with the attractive Kelbg potential (grey). The Drude correction is insufficient to capture the peak at the plasma frequency. MD simulations with the repulsive Kelbg potential (black) show that free electron motion still contributes to a peak at the plasma frequency.}
\end{figure}

The high and low frequency models from Secs.~\ref{subsec:HF} and~\ref{subsec:LF} can be combined to form one comprehensive solution. This is done by multiplying the high frequency behavior onto the low frequency Drude solution such that
\begin{eqnarray}
    \label{eq:comprehensive}
    j_{D}(\omega) = j_{l}(\omega) \frac{j_{h}(\omega)}{j_{h}(\omega\xrightarrow{}0)}.
\end{eqnarray}
This comprehensive solution compares well with the MD data, as shown in Fig.~\ref{RepulsiveandAttractive}. However, at strong coupling there is a peak near the plasma frequency, as shown in the case of $\Gamma = 1$ and $\Theta = 0.468$ in Fig.~\ref{RepulsiveandAttractive}. Here, the Drude form for the low-frequency part of the comprehensive solution is insufficient to describe this feature. Our previous work~\cite{kinney2024mean} demonstrated that it is possible to capture this peak in the repulsive case by modeling the oscillations in the correlation functions at strong coupling. The attractive case is more complicated because there will be some contribution of bound electrons to this peak. However, the peak is still visible in the spectrum generated from repulsive interactions, where bound states are not possible. This suggests that free electronic motion contributes to the peak as well. 

\section{\label{sec:discussion}Discussion}
Historically, the frequency dependence of bremsstrahlung emission is captured using the Gaunt factor.\cite{gaunt1930continuous,johnston1967free} The emission coefficient in hydrogen ($Z = 1$) is related to the Gaunt factor $G(\omega)$ as
\begin{eqnarray}
\label{eq:GauntFactorandEmissionCoefficient}
    j(\omega) = \frac{n_{e}n_{i}q^{6}}{12\pi^{3}\epsilon_{0}^{3}c^{3}m_{e}^{2}}\sqrt{\frac{\pi m_{e}}{6k_{B}T}}G(\omega). 
\end{eqnarray}
Traditional models for the Gaunt factor assume that the plasma is weakly coupled.~\cite{landau1971classical,oster1961emission,befeki1966radiation,dawson1962high} Section~\ref{subsec:TradModels} compares the Gaunt factor obtained from mean force emission theory with these classical formulas and shows that there is better agreement with the new method when both screening and quantum effects are included. However, the classical Gaunt factors are still limited to weak coupling and give negative values at high frequencies.
%Section~\ref{subsec:TradModels} compares the Gaunt factor obtained from mean force emission theory with these classical formulas and shows that the new method is able to capture screening and quantum effects that are absent in the traditional formulas.
Sections~\ref{subsec:ConservationofEnergy} and ~\ref{subsec:FermiDirac} address how the conservation of energy and including a Fermi-Dirac distribution of electrons would be expected to impact the bremsstrahlung emission. Finally,
Sec.~\ref{subsec:Limitations} compares mean force emission theory with the Born approximation of Sommerfeld's quantum result~\cite{sommerfeld1924atombau} and discusses a limitation of the Kelbg potential approach at capturing very high frequency emission. However, the potential of mean force can capture the impact of screening. A calculation of the total bremsstrahlung power is given to show that screening leads to a reduction in the total bremsstrahlung power, even in a weakly coupled plasma. 

\subsection{\label{subsec:TradModels}Comparison with Traditional Classical Gaunt Factors}
\begin{figure}[t!]\label{weaklycoupledg0p01}
\includegraphics[width=0.5\textwidth]{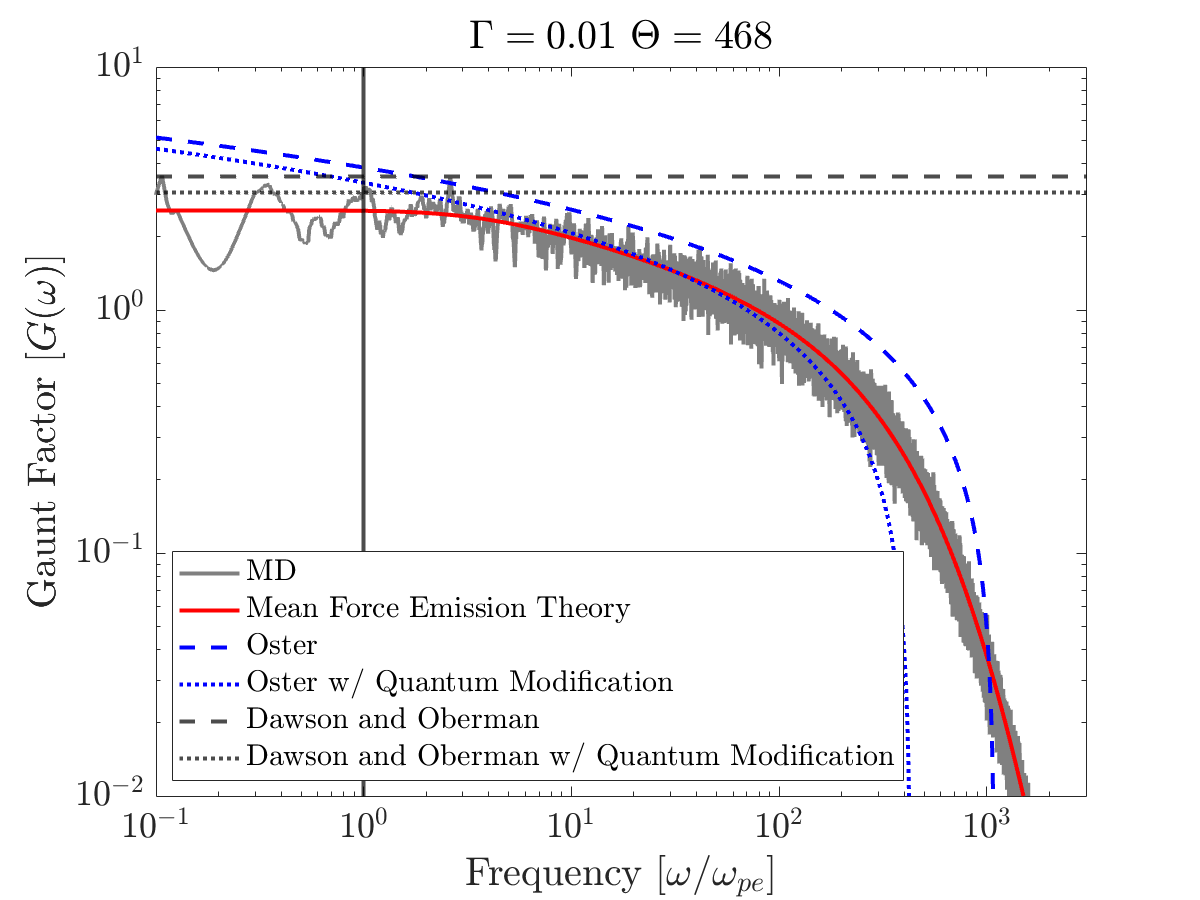}
    \caption{\label{weaklycoupledg0p01} Gaunt factor calculated using traditional classical models compared to mean force emission theory at weak coupling. The curves based on Oster's model use Eq.~(\ref{eq:GauntFactorClassicalCoulombLowFreq}) (dashed blue) and Eq.~(\ref{eq:GauntFactorClassicalCoulombLowFreqQuantumMod}) (dotted blue). The Dawson and Oberman result in the $\omega\ll\omega_{pe}$ limit is obtained from Eq.~(\ref{eq:dawsonandoberman}) (horizontal dashed black line) and the modified result is based on Eq.~(\ref{eq:dawsonandobermanquant}) (horizontal dotted black line).}
\end{figure}

The traditional form of a Gaunt factor is given by\cite{befeki1966radiation}
\begin{eqnarray}
\label{eq:ClassicalGauntFactor}
    G(\omega) = \sqrt{\frac{8\pi^{3} k_{B}T}{m_{e}}} \frac{1}{n_e}\int_{0}^{\infty}dv\, vf_{M}(v)g(\omega,v),
\end{eqnarray}
where $g(\omega,v)$ is some velocity-dependent term that has a different form in different models. The most common formula is Oster's model~\cite{oster1961emission,landau1971classical,befeki1966radiation}. Here, the electron trajectory is computed in the same way as in the high frequency limit discussed in Sec.~\ref{subsec:HF}, but where the interaction is modeled using the attractive Coulomb force. This leads to the velocity-dependent Gaunt factor  
\begin{eqnarray}
    \label{eq:GauntFactorClassicalCoulomb}
    g(\omega,v) = \frac{\sqrt{3}}{\pi}\Omega|K_{i\Omega}(\Omega)||K_{i\Omega}^{'}(\Omega)|e^{\pi\Omega},
\end{eqnarray}
where $\Omega = \omega b_{\min}/v$, $b_{\min} = Zq^{2}/4\pi\epsilon_{0}m_{e}v^{2}$ is the classical distance of closest approach for an electron with initial speed $v$, and $K_{i\Omega}(\Omega)$ and $K^{'}_{i\Omega}(\Omega)$ represent the modified Hankel function of order $i\Omega$ and its derivative. In the high frequency limit, the Hankel function and its derivative are expanded for $\Omega \gg 1$ and give $G(\omega) = 1$, which means the emission coefficient is frequency-independent as $\omega\xrightarrow{}\infty$.~\cite{landau1971classical} This result is a direct consequence of the $\sim1/r$ behavior of the attractive Coulomb potential. The divergence of the potential at short distances causes arbitrarily high forces on electrons that are close to ions and leads to radiation at arbitrarily high frequencies. The Kelbg potential does not have this asymptotic behavior because the force on an electron at close distances from an ion is finite.

In the low frequency limit, the Hankel function and its derivative are expanded for $\Omega \ll 1$ and give the following result for the Gaunt factor\cite{oster1961emission,landau1971classical,befeki1966radiation}
\begin{eqnarray} 
    \label{eq:GauntFactorClassicalCoulombLowFreq}
    G_\textrm{Oster}(\omega) = \frac{\sqrt{3}}{\pi}\ln\biggl(\frac{4}{e^{5\gamma/2}}\frac{v_{Te}}{r_{L}} \frac{1}{\omega}\biggr),
\end{eqnarray}
where $\gamma \approx 0.577$ is Euler's constant. This is identical to the repulsive Coulomb case and features the frequency-dependent `Coulomb logarithm'. This result is only valid in the limit $\Omega \ll 1$, but takes a non-physical negative value as $\omega\xrightarrow{} v_{Te}/r_{L}$. This formula also diverges as $\omega\xrightarrow{}0$ because the Coulomb potential neglects plasma screening, as shown in Fig.~\ref{weaklycoupledg0p01}. 

A correction to Eq.~(\ref{eq:GauntFactorClassicalCoulombLowFreq}) to include screening was developed by Dawson and Oberman using a linear-dielectric approach.\cite{dawson1962high} In the $\omega\gg\omega_{pe}$ limit, the result is identical to Eq.~(\ref{eq:GauntFactorClassicalCoulombLowFreq}). In the $\omega\ll\omega_{pe}$ limit, the Gaunt factor takes a constant value~\cite{befeki1966radiation}
\begin{eqnarray}
    \label{eq:dawsonandoberman}
    G_\textrm{DO}(\omega) = \frac{\sqrt{3}}{\pi}\ln\biggl(\frac{2^{3/2}}{e^{2\gamma+1/2}}\frac{v_{Te}}{\omega_{pe} r_{L}}\biggr), 
\end{eqnarray}
a result that qualitatively agrees with mean force emission theory and is a result of the linear-dielectric approach including plasma screening. Mean force emission theory naturally includes this low frequency plateau, but also generalizes it to strong coupling as a result of modeling electron-ion interactions with the potential of mean force. 

The two analytic results from Eqs.~(\ref{eq:GauntFactorClassicalCoulombLowFreq}) and (\ref{eq:dawsonandoberman}) systematically over-predict the Gaunt factor in comparison with the MD calculations (see Fig.~\ref{weaklycoupledg0p01}). One reason for this is the neglect of the quantum-mechanical nature of the electron-ion interaction at close distances, which is necessary when the Kelbg scale length exceeds the distance of closest approach in a classical interaction, $\lambda > r_\textrm{L}$. For $Z = 1$, this translates to the condition that the temperature $T\gtrsim 55 \textrm{eV}$. Accounting for the quantum cutoff at close distances, Oster's formula from Eq.~(\ref{eq:GauntFactorClassicalCoulombLowFreq}) is modified to~\cite{oster1961emission,befeki1966radiation}
\begin{eqnarray} 
    \label{eq:GauntFactorClassicalCoulombLowFreqQuantumMod}
    G_\textrm{Oster,q}(\omega) = \frac{\sqrt{3}}{\pi}\ln\biggl(\frac{2}{e^{\gamma}}\frac{v_{Te}}{\omega\lambda}\biggr).
\end{eqnarray}
Besides a numerical factor of order unity in the logarithm, the quantum modifications to the classical Gaunt factor simply replaces the Landau length with the scale length for the Kelbg potential. Note that this result is identical to the low frequency limit of the Gaunt Factor in the Born approximation; see Sec.~\ref{subsec:Limitations}. The low frequency Dawson and Oberman result has also been modified in a similar way, leading to~\cite{befeki1966radiation}
\begin{eqnarray}
    \label{eq:dawsonandobermanquant}
    G_\textrm{DO,q}(\omega) = \frac{\sqrt{3}}{\pi}\ln\biggl(\sqrt{\frac{2}{e^{\gamma+1}}}\frac{v_{Te}}{\omega_{pe} \lambda}\biggr). 
\end{eqnarray}
Figure~\ref{weaklycoupledg0p01} shows that these simple replacements improve the agreement with mean force emission theory somewhat, particularly near the plasma frequency. Although this physics is included as an ad hoc correction in the classical theories, the MD simulations and mean force emission theory capture it through the Kelbg potential.

Finally, the Appendix shows that it is possible to connect the limits of Oster's Gaunt factor in Eqns.~(\ref{eq:GauntFactorClassicalCoulombLowFreq}) and (\ref{eq:GauntFactorClassicalCoulombLowFreqQuantumMod}) with the low frequency Dawson and Oberman formulas in Eqns.~(\ref{eq:dawsonandoberman}) and (\ref{eq:dawsonandobermanquant}) by reframing the problem in terms of dynamic structure factors. This results in a simple analytic formula that corrects Oster's Gaunt factor for screening.

\subsection{\label{subsec:ConservationofEnergy}Conservation of Energy}
An important aspect of the bremsstrahlung spectrum is the exponential decay at high frequencies.~\cite{befeki1966radiation} This feature is a result of conservation of energy - a photon of energy $\hbar\omega$ cannot be emitted by an electron with less kinetic energy $(1/2)m_{e}v^{2} < \hbar\omega$. A purely classical approach cannot capture this.

Generally there are two ways this physics is included in the literature. The first is to restrict the velocity averaging in the calculation of the classical Gaunt factor by placing a lower bound $v_{\min} = \sqrt{2\hbar\omega/m_{e}}$ on the integral in Eq.~(\ref{eq:ClassicalGauntFactor}).~\cite{befeki1966radiation} The reasoning behind this ad-hoc modification is that an electron needs some minimum kinetic energy to emit a photon. The second, and more rigorous approach, is to retain the effects of detailed balance (the exponential factor) in the calculation of the dynamic conductivity given by Eq.~(\ref{eq:dynamicconductivityFACF1}). This amounts to not taking the limit $\hbar\omega\ll k_{B}T$. Then, taking the current operators to be classical variables in Eq.~(\ref{eq:dynamicconductivityFACF1}) and applying the relations in Sec.~\ref{subsec:RelationtoAutocorrelationFunctions} with the full blackbody spectrum in Eq.~(\ref{eq:blackbody}). Doing so leads to
\begin{eqnarray}
\label{eq:emissioncoefficientFACFexp}
    j(\omega) = \frac{e^{-\frac{\hbar\omega}{k_{B}T}}}{6\pi^2\epsilon_{0}c^{3}V}\mathrm{Re}\biggl\{\int_{0}^{\infty}\langle\dot{\mathbf{J}}(t)\cdot\dot{\mathbf{J}}(0)\rangle e^{-i\omega t}dt\biggr\},
\end{eqnarray}
which predicts a characteristic decay of the spectrum when $\hbar\omega/(k_{B}T) \gtrsim 1$, or $\omega_\textrm{decay}/\omega_{pe}\approx0.88\sqrt{\Theta/\Gamma}$ for $Z = 1$. Comparison with the characteristic decay due to the Kelbg potential (Eq.~(\ref{eq:omega_max_kelbg})) shows that the decay occurs at approximately the same frequency in both models, but is always at a slightly lower frequency when the exponential factor is retained. 

\subsection{\label{subsec:FermiDirac}Fermi Dirac Distribution}
\begin{figure}[t!]\label{fermidirac}
\includegraphics[width=0.5\textwidth]{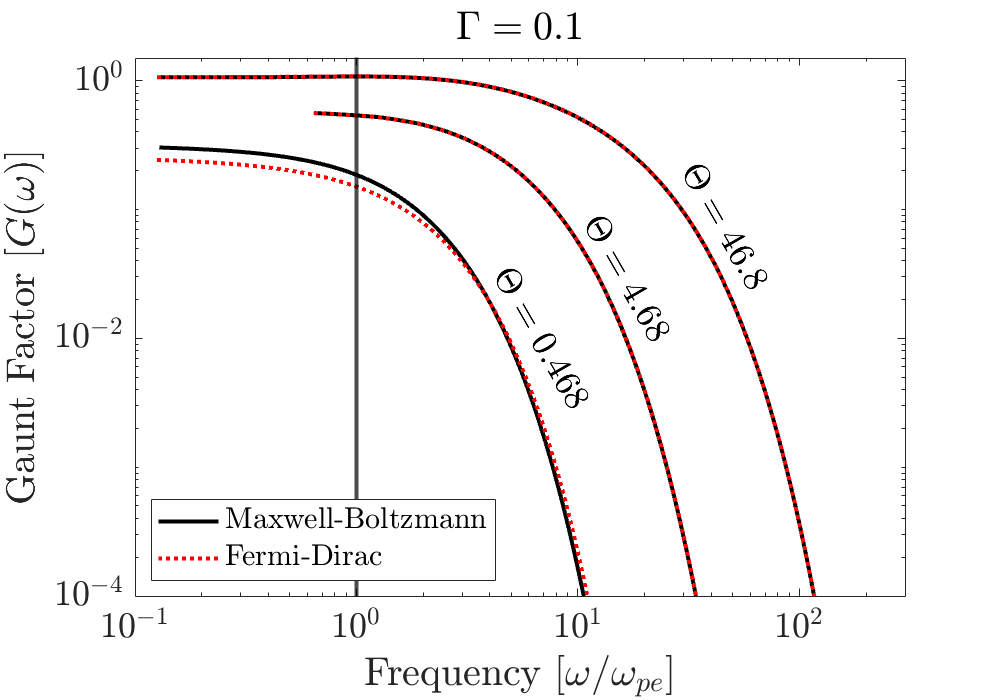}
    \caption{\label{fermidirac} Integration over a Fermi-Dirac distribution at degenerate conditions ($\Theta < 1$) enhances the high frequency emission while decreasing the low frequency plateau. The chemical potential is determined for a free-electron gas. All curves include the exponential factor discussed in Sec.~\ref{subsec:ConservationofEnergy}.} 
\end{figure}

The classical MD simulations used in this paper cannot capture the Fermi-Dirac distribution of electrons that forms in a degenerate plasma. To assess the impact of this on high frequency bremsstrahlung emission,  the velocity-dependent Gaunt factor is integrated over a Fermi-Dirac distribution, 
\begin{eqnarray}
    \label{eq:FD}
    f_\textrm{FD}(v) = \frac{m_{e}^{3}}{4\pi^{3}\hbar^{3}}\frac{1}{\exp{\left(v^{2}/v_{Te}^{2}-\eta\right)}+1},
\end{eqnarray}
such that
\begin{eqnarray}
    \label{eq:GauntFactorFD}
    G_\textrm{FD}(\omega) = \sqrt{\frac{8\pi^{3} k_\textrm{B}T}{m_{e}}}e^{-\frac{\hbar\omega}{k_\textrm{B}T}}\int_{0}^{\infty}dv\, v \frac{f_\textrm{FD}(v)}{n_{e}}g(\omega,v).
\end{eqnarray}
The normalized chemical potential $\eta$ is determined for a free-electron gas by the condition $I_{1/2}(\eta) = (2/3)\Theta^{-3/2}$, where
\begin{eqnarray}
    \label{eq:FermiIntegral}
    I_{1/2}(\eta) = \int_{0}^{\infty}dt\frac{\sqrt{t}}{e^{t-\eta}+1}
\end{eqnarray}
is the Fermi integral. 
A convenient fit relating $\eta$ and $\Theta$ is provided by Ichimaru~\cite{ichimaru2004statistical} 
\begin{eqnarray}
    \label{eq:IchimaruFit}
    \eta = -\frac{3}{2}\ln{\left(\Theta\right)} + \ln{\left(\frac{4}{3\sqrt{\pi}}\right)} + \frac{A\Theta^{-(b+1)}+B\Theta^{-(b+1)/2}}{1+A\Theta^{-b}},
\end{eqnarray}
where $A = 0.25954$, $B = 0.072$, and $b = 0.858$. This expression is accurate to within $0.2\%$ for all values of $\Theta$.

For $\Theta\leq1$, degeneracy reduces the population of low speed electrons, while increasing the population of high speed electrons. Low speed electrons primarily contribute to lower frequency emission near the plasma frequency while high speed electrons have more of an impact on the very high frequency part of the emission spectrum. Thus, the effect of the degenerate electron distribution is to lower emission for $\omega\sim\omega_{pe}$ and increase the high frequency emission (see Fig.~\ref{fermidirac}). The emission spectrum might also be impacted because the degenerate electron distribution will also change the potential of mean force. However, this is expected to be a small effect for $\Theta\sim1$.

\begin{figure}[h!]\label{borncompare}
\includegraphics[width=0.5\textwidth]{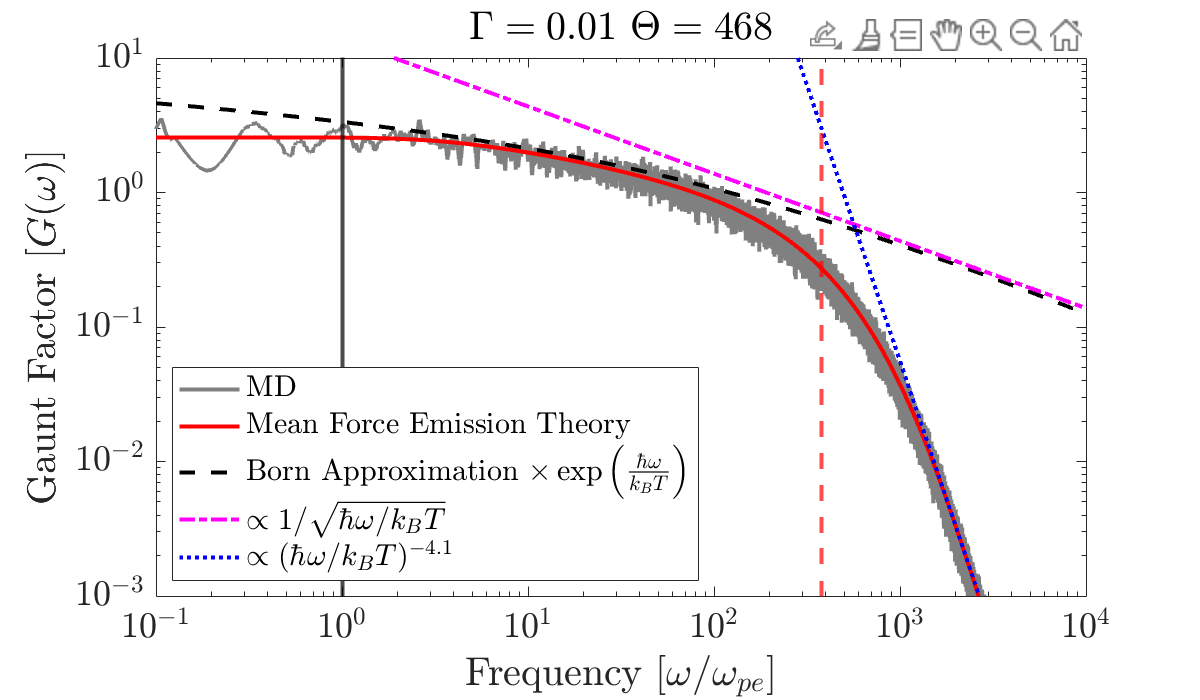}
    \caption{\label{borncompare} The Gaunt factor from MD, mean force emission theory, and the Born approximation  from Eq.~(\ref{eq:born}) times $\exp(\hbar\omega/k_BT)$ are all plotted without the exponential decay factor discussed in Sec.~\ref{subsec:ConservationofEnergy}. The red-dashed vertical line denotes the characteristic decay due to Kelbg potential from Eq.~(\ref{eq:omega_max_kelbg}).} 
\end{figure}

\subsection{\label{subsec:Limitations}Limitations of the Kelbg Potential and Impact of Screening on the Total Bremsstrahlung Power}
 %First, we illustrate the limitations of using the Kelbg potential to determine very high frequency results by comparing 
%Here, we compare mean force emission theory with the Born approximation of Sommerfeld's seminal result.\cite{sommerfeld1924atombau} Other work has found that while collision frequencies (and thus the low frequency spectrum) obtained from the Kelbg potential are in reasonable agreement with experimental data, the high frequency limit of the spectrum is not consistent with quantum mechanical calculations.~\cite{morozov2005molecular} Then, we use a model that combines the Born approximation and mean force emission theory to show that screening reduces the total bremsstrahlung power.

Sommerfeld's original work is based on a quantum mechanical solution to electron-ion scattering through a Coulomb potential, analogous to the classical high frequency limit given in Sec.~\ref{subsec:TradModels}. The exact form is difficult to evaluate as it is written in terms of a hypergeometric function, but in the limit of high incoming electron speeds (Born approximation) the Gaunt factor is
\begin{eqnarray}
    \label{eq:born}
    G(\omega) = \frac{\sqrt{3}}{\pi}e^{-\frac{\hbar\omega}{2k_{B}T}}K_{0}\left(\frac{\hbar\omega}{2k_{B}T}\right)
\end{eqnarray}
where $K_{0}$ is the modified Hankel function of order zero and the exponential decay factor $\exp(-\hbar\omega/k_{B}T)$ is implicitly included. 
Note that part of the usual exponential decay term is in the $K_0$ function, leading to the factor of $\exp(-\hbar\omega/2k_{B}T)$. 
Figure~\ref{borncompare} plots the Born approximation and mean force emission theory without the exponential decay (obtained by multiplying Eq.~(\ref{eq:born}) by $\exp(\hbar \omega/k_BT)$).

\begin{figure}[t!]\label{piecewisemodel}
\includegraphics[width=0.5\textwidth]{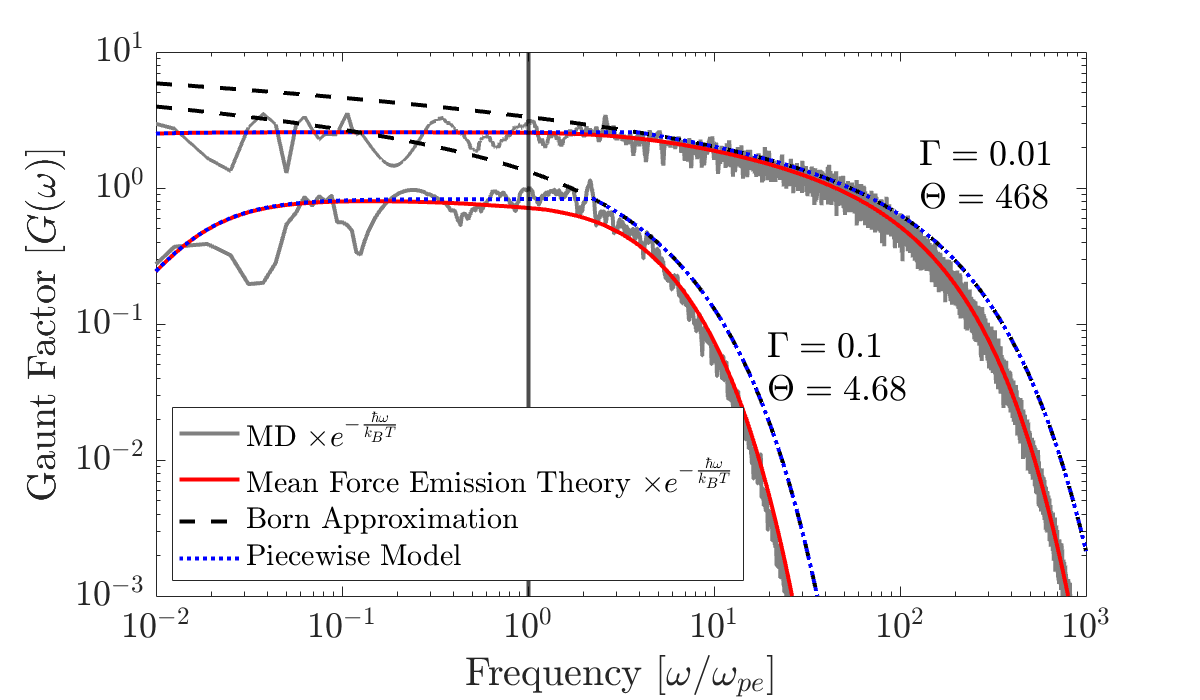}
    \caption{\label{piecewisemodel} The piecewise model given by Eq.~(\ref{eq:BornplusMFET}) combines the high frequency Born approximation result with the low frequency plateau from mean force emission theory.} 
\end{figure}

\begin{figure}[t!]\label{power}
\includegraphics[width=0.5\textwidth]{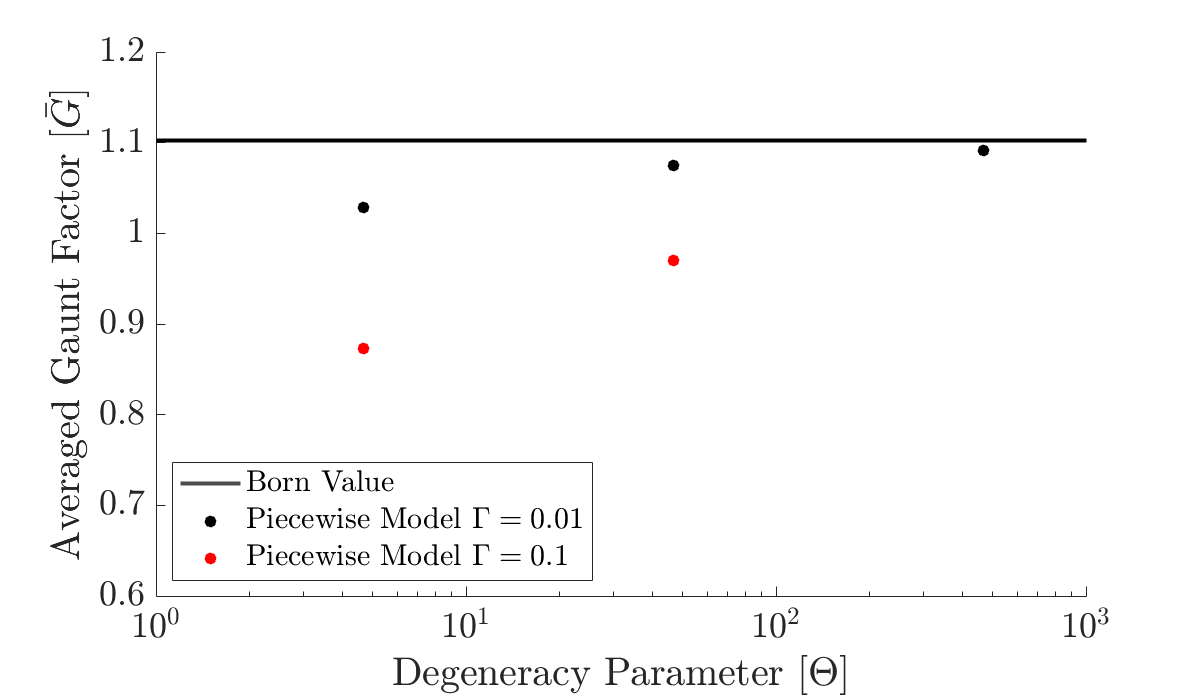}
    \caption{\label{power} Frequency averaged Gaunt factor $\Bar{G}$ calculated for the weakly coupled $\Gamma < 1$ and non-degenerate $\Theta > 1$ MD simulation conditions explored in this paper. The calculations use the piecewise model in Eq.~(\ref{eq:BornplusMFET}).} 
\end{figure}

For frequencies $\omega>\omega_\textrm{max,K}$ there is a clear deviation between mean force emission theory and the Born approximation. Other work has found that while collision frequencies (and thus the low frequency spectrum) obtained from the Kelbg potential are in reasonable agreement with experimental data, the high frequency limit of the spectrum is not consistent with quantum mechanical calculations.~\cite{morozov2005molecular} At high frequencies, mean force emission theory follows a decay $\propto (\hbar\omega/k_{B}T)^{-4.1}$. Previous work found power law behavior $\propto\omega^{-(3.4-3.8)}$ for simulations where $\Gamma = 1$.~\cite{morozov2005molecular} The Born approximation also shows a decay at high frequencies, which suggests that the decay of the spectrum due to the quantum effects is at least qualitatively captured by the Kelbg interaction. Although, the Born approximation formula follows a decay $\propto (\hbar\omega/k_{B}T)^{-0.5}$. However, the result of the Born approximation here should not be taken as the right answer and further work will need to compare with exact quantum mechanical calculations.

Importantly, the Born approximation formula in Eq.~(\ref{eq:born}) does not capture screening because it is based on electron-ion scattering through the bare Coulomb potential. Mean force emission theory does include screening and is thus able to capture the plateau that occurs as $\omega\xrightarrow{}\omega_{pe}$. This plateau effect due to screening will reduce the area under the emission spectrum and the total bremsstrahlung power. The radiated power density per sterradian ($P$) can be calculated by integrating the emission coefficient over all frequencies such that, for $Z = 1$   
\begin{eqnarray}
    \label{eq:TotalPower}
    P = \frac{n_{e}n_{i}q^{6}}{48\pi^{4}\epsilon_{0}^{3}\hbar m_{e}c^{2}}\sqrt{\frac{2\pi k_{B}T}{3m_{e}c^{2}}}\Bar{G}
\end{eqnarray}
where 
\begin{eqnarray}
    \label{eq:AvgGauntFactor}
    \Bar{G} = \int_{0}^{\infty}G\left(\frac{\hbar\omega}{k_{B}T}\right)d\left(\frac{\hbar\omega}{k_{B}T}\right)
\end{eqnarray}
is the frequency averaged Gaunt factor. For the Gaunt factor in the Born approximation, $\Bar{G} = 2\sqrt{3}/\pi\approx 1.1$. Using $\Bar{G}\approx1.1$ in Eq.~(\ref{eq:TotalPower}) gives the result that is used in the NRL Plasma Formulary.~\cite{huba1998nrl}

In order to assess the impact of screening for $\Gamma < 1$ and $\Theta > 1$ on the total bremsstrahlung power, we take the Born result at high frequencies and patch it together with the Gaunt factor from the low frequency Drude formula given in Eq.~(\ref{eq:LFemissioncoefficient}) (see Fig.~\ref{piecewisemodel}). This gives the following piecewise function
\begin{eqnarray}
\label{eq:BornplusMFET}
G(\omega) = \begin{cases}
\frac{3\nu_{ei}}{\sqrt{\pi}\Gamma^{3/2}\omega_{pe}}\frac{\omega^{2}}{\nu_{ei}^{2}+\omega^{2}} & \text{if } \omega < x \\
\frac{\sqrt{3}}{\pi}e^{-\frac{\hbar\omega}{2k_{B}T}}K_{0}\left(\frac{\hbar\omega}{2k_{B}T}\right) & \text{if } \omega \ge x
\end{cases}
\end{eqnarray}
where $x$ is the point in frequency space connecting the two solutions. For plasma conditions that are very weakly coupled $\Gamma \ll 1$ and non-degenerate $\Theta \gg 1$, screening will have less of an impact on the total power (see Fig.~\ref{power}). As $\Gamma,\Theta\xrightarrow{}1$, screening in the weakly coupled regime will slightly lower the total bremsstrahlung power. %However, the traditional Born approximation result for $\Bar{G}$ is quite good over a range of weakly coupled and non-degenerate conditions.

\section{\label{sec:conclusion}Conclusion}
Mean force emission theory was developed in order to extend classical bremsstrahlung theory to the strongly coupled regime. Through comparison with classical MD simulations, this paper shows that the same framework can be applied to describe electron-ion plasmas. The attractive electron-ion interaction introduces several differences from the previously calculated spectra in the positron-ion case.~\cite{kinney2024mean} For one, the attractive interaction modifies the high frequency spectrum by changing the time-history of the electron force during a collision with a single ion. The attractive interaction also changes the electron-ion collision frequency and thus the low frequency part of the spectrum. Furthermore, the attractive interaction allows for classically bound electron-ion states to form. These bound states create peaks in the emission spectrum associated with the electron orbits. 

In order to avoid Coulomb collapse, electron-ion interactions were modeled using the Kelbg potential. The specific form of the Kelbg potential also introduces several qualitative features. For one, there is a high frequency decay associated with the softening of the electron-ion force at the thermal deBroglie wavelength. The Kelbg potential also changes the electron-ion collision frequency by lowering the scattering angle during close electron-ion encounters. It is shown that the Gaunt factor can be extracted from the simulations, and comparison with traditional models suggests that the Kelbg potential is able to capture ad-hoc quantum modifications to the classical results. We also discuss a limitation of the Kelbg potential for studying bremsstrahlung emission at very high frequencies and find similar results to previous work.\cite{morozov2005molecular}

Importantly, this paper still supports the conclusion found earlier in the repulsive system, that the classical high frequency solution with the Drude correction cannot capture a peak at the plasma frequency that forms at strong coupling. This peak is argued to be associated with the strongly correlated motion of free electrons in the plasma - independent of the peaks due to classically bound particles. Furthermore, we find that plasma screening leads to a plateau in the spectrum near the electron plasma frequency, and find that this reduces the total bremsstrahlung power.

Finally, the calculation of the classical bremsstrahlung spectrum from a force autocorrelation function is also intimately related to other transport quantities such as absorption and the real part of the dynamic conductivity. Therefore, this paper also fleshes out a powerful technique to explore these transport coefficients. A fully quantum mechanical calculation of the emission, absorption, and conductivity in the linear response regime all come from the same Kubo equation. The framework here thus provides a useful foundation for exploring fully quantum-mechanical descriptions in the future.

\begin{acknowledgments}
The authors thank Lucas J. Babati for helpful conversations on this work. This work is funded by the NNSA Stockpile Stewardship Academic Alliances under grant number DE-NA0004100 and the DOE NNSA Stockpile Stewardship Graduate Fellowship through cooperative agreement DE-NA0003960. Additionally, this research was supported in part through computational resources and services provided by Advanced Research Computing (ARC), a division of Information and Technology Services (ITS) at the University of Michigan, Ann Arbor.
\end{acknowledgments}

\section*{Data Availability Statement}
The data that support the findings of this study are available from the corresponding author upon reasonable request.

%\section{Appendix}
\appendix
\section{Connection with the dynamic structure factor}
Here we show that the general framework for bremsstrahlung given by Eq.~(\ref{eq:emissioncoefficientFACF}) can be written in terms of the electron-electron and ion-ion dynamic structure factor. This result was earlier used in a paper by Iglesias~\cite{iglesias1996corrections} and is similar to the expression obtained by Ichimaru~\cite{ichimaru2004statistical}. We begin by noting that for a two component plasma of electrons and ions one may expand in terms of the mass ratio $m_{i}\gg m_{e}$ and see that the forces between electrons cancel out so that
\begin{align}
\label{eq:ignore_ions_and_electron_forces}
    \nonumber\dot{\mathbf{J}}(t) = \sum_{j}\frac{q_{j}}{m_{j}}\mathbf{F}_{j}(t) & \approx-\frac{q}{m_{e}}\sum_{m=1}^{N_{e}}\mathbf{F}_{m}(t)\\
    &\approx -\frac{q}{m_{e}}\sum_{m=1}^{N_{e}}\sum_{n=1}^{N_{i}}\mathbf{F}_{m,n}(t)
\end{align}
where $\mathbf{F}_{m,n}(t)$ is the force on electron $m$ due to ion $n$. The total force on electron $m$ due to all the ions is
\begin{align}
\label{eq:spatialFT}
    \nonumber\mathbf{F}_{m}(t) &= -\frac{\partial}{\partial\mathbf{r}_{m}}\sum_{n=1}^{N_{i}}\frac{(Zq)(-q)}{4\pi\epsilon_{0}|\mathbf{r}_{m}-\mathbf{r}_{n}|}\\
    &= \frac{i}{(2\pi)^{3}}\sum_{n=1}^{N_{i}}\int d^{3}k\mathbf{k}V(k)e^{-i\mathbf{k}\cdot(\mathbf{r}_{n}-\mathbf{r}_{m})}
\end{align}
where $V(k) = Zq^{2}/(k^{2}\epsilon_{0})$ is the Fourier transform of the Coulomb potential. Then the spatial Fourier transform of the species $j$ density is defined as $n_{j}(\mathbf{k},t)=\sum_{j}\exp(i\mathbf{k}\cdot\mathbf{r}_{j})$ so that 
\begin{eqnarray}
\label{eq:spatialFT2}
    \sum_{m=1}^{N_{e}}\mathbf{F}_{m}(t) = \frac{i}{(2\pi)^{3}}\int d^{3}kV(k)\mathbf{k}n_{e}(\mathbf{k},t)n_{i}^{*}(\mathbf{k},t).
\end{eqnarray}
Substitution into Eq.~(\ref{eq:emissioncoefficientFACF}) gives
\begin{eqnarray}
\label{eq:full}
    \nonumber j(\omega) =\frac{q^{2}}{96\pi^{5}\epsilon_{0}c^{3}V^{2}m_{e}^{2}}\int d^{3}k\left[kV(k)\right]^{2}\\
    \times\int_{-\infty}^{\infty}dte^{-i\omega t} \langle Q(\mathbf{k},t)\rangle
\end{eqnarray}
where 
\begin{eqnarray}
    \label{eq:Qkt}
    \langle Q(\mathbf{k},t)\rangle = \langle n_{e}^{*}(\mathbf{k},t)n_{i}(\mathbf{k},t)n_{e}(\mathbf{k},0)n_{i}^{*}(\mathbf{k},0)\rangle
\end{eqnarray}
is a quadruple correlation. Under the assumption that the electron and ion density variations are decoupled, Eq.~(\ref{eq:Qkt}) becomes
\begin{eqnarray}
    \label{eq:Qktapprox}
    \nonumber\langle Q(\mathbf{k},t)\rangle\approx\langle n_{e}^{*}(\mathbf{k},t)n_{e}(\mathbf{k},0)\rangle\langle n_{i}(\mathbf{k},t)n_{i}^{*}(\mathbf{k},0)\rangle\\
    = N_{e}F_{ee}(-\mathbf{k},t)N_{i}F_{ii}(\mathbf{k},t)
\end{eqnarray}
where $F_{jj} = \frac{1}{N_{j}}\langle n_{j}(\mathbf{k},t)n_{j}^{*}(\mathbf{k},0)\rangle$ is the intermediate scattering function. Using the convolution theorem, the temporal Fourier transform of Eq.~(\ref{eq:Qktapprox}) can be written as
\begin{eqnarray}
    \label{eq:SF1}
    \nonumber N_{e}N_{i}\int_{-\infty}^{\infty}dte^{-i\omega t}F_{ee}(-\mathbf{k},t)F_{ii}(\mathbf{k},t)\\
    = 2\pi N_{e}N_{i}\int_{-\infty}^{\infty}d\omega'S_{ee}(-\mathbf{k},\omega')S_{ii}(\mathbf{k},\omega-\omega')
\end{eqnarray}
where $S_{jj}(\mathbf{k},\omega)$ is the dynamic structure factor defined as the Fourier transform of the intermediate scattering function. Applying the commutativity of the convolution, the bremsstrahlung emission coefficient is
\begin{eqnarray}
\label{eq:emissioncoefficientSF}
    \nonumber j(\omega) \approx \frac{q^{2}n_{i}n_{e}}{48\pi^4\epsilon_{0}c^{3}m_{e}^{2}}\int d^{3}k\left[kV(k)\right]^{2}\\
    \times\int_{-\infty}^{\infty}d\omega'S_{ee}(-\mathbf{k},\omega-\omega')S_{ii}(\mathbf{k},\omega').
\end{eqnarray}
Finally, at high frequencies the ions are considered to be static so that $S_{ii}(\mathbf{k},\omega^{'}) = S_{ii}(\mathbf{k})\delta(\omega')$. Then
the emission coefficient is
\begin{eqnarray}
\label{eq:emissioncoefficientSF2}
    \nonumber &j(\omega) \approx \frac{q^{2}n_{i}n_{e}}{48\pi^4\epsilon_{0}c^{3}m_{e}^{2}}\\
    &\times\int d^{3}k\left[kV(k)\right]^{2}S_{ee}(-\mathbf{k},\omega)S_{ii}(\mathbf{k})
\end{eqnarray}
and the Gaunt factor is
\begin{eqnarray}
\label{eq:GauntFactorSF}
    \nonumber &G(\omega) \approx \sqrt{\frac{3k_{B}T}{8\pi^{3}m_{e}}}\frac{\epsilon_{0}^{2}}{Z^2q^4}\\
    &\times\int d^{3}k\left[kV(k)\right]^{2}S_{ee}(-\mathbf{k},\omega)S_{ii}(\mathbf{k}).
\end{eqnarray}
Equation~(\ref{eq:emissioncoefficientSF}) is similar to the result of Ichimaru~\cite{ichimaru2004statistical}, except that his result includes the electron-ion and ion-electron dynamic structure factors. Equation~(\ref{eq:GauntFactorSF}) is identical to the expression used by Iglesias.\cite{iglesias1996corrections}

\section{Analytic screening corrections to Oster's Gaunt factor in a weakly coupled and non-degenerate plasma}
Starting with Eq.~(\ref{eq:GauntFactorSF}), one can derive Oster's classical (Eq.~(\ref{eq:GauntFactorClassicalCoulombLowFreq})) and quantum mechanically modified (Eq.~(\ref{eq:GauntFactorClassicalCoulombLowFreqQuantumMod})) Gaunt factors by assuming forms of the ion-ion and electron-electron dynamic structure factors. The ion positions are assumed to be uncorrelated ($S_{ii}(k) \approx 1$) and the electrons are approximated as an ideal gas so that
\begin{eqnarray}
    \label{eq:eeSkomega}
    S_{ee}(-\mathbf{k},\omega)\approx \sqrt{\frac{m_{e}}{2\pi k^{2}k_{B}T}}\exp{\left(-\frac{m_{e}\omega^{2}}{2k^{2}k_{B}T}\right)}.
\end{eqnarray}
Then the Gaunt factor becomes
\begin{eqnarray}
\label{eq:GauntFactorSFOster1}
    G(\omega) \approx \frac{\sqrt{3}}{\pi}\int_{0}^{k_{\max}}\frac{dk}{k}\exp{\left(-\frac{k_{\min}^{2}}{k^{2}}\right)}
\end{eqnarray}
where $k_{\min} = \omega/v_{Te}$. The upper cutoff ($k_{\max}$) on the integral is necessary because the ideal gas form of the electron-electron structure factor does not properly account for close interactions. Performing the integral and taking the limit $k_{\min}/k_{\max}\ll1$, the result is
\begin{eqnarray}
\label{eq:GauntFactorSFOster2}
    G(\omega) \approx \frac{\sqrt{3}}{\pi}\ln{\left(\frac{1}{e^{\gamma/2}}\frac{k_{\max}}{k_{\min}}\right)}.
\end{eqnarray}
Using $k_{\max} = 4/(e^{2\gamma}r_{L})$ gives the Oster's classical Gaunt factor in (Eq.~(\ref{eq:GauntFactorClassicalCoulombLowFreq})). Using $k_{\max} = 2/(e^{\gamma/2}\lambda)$ gives the Oster's quantum mechanically modified Gaunt factor in (Eq.~(\ref{eq:GauntFactorClassicalCoulombLowFreqQuantumMod})). 

A simple formula that accounts for static screening in Oster's results can be derived by adjusting the electron-electron structure factor~\cite{ichimaru2004statistical} such that 
\begin{eqnarray}
    \label{eq:eeSkomega}
    S_{ee}(-\mathbf{k},\omega)\approx \frac{1}{\lvert\epsilon(-\mathbf{k},\omega)\rvert^{2}}\sqrt{\frac{m_{e}}{2\pi k^{2}k_{B}T}}e^{-\frac{m_{e}\omega^{2}}{2k^{2}k_{B}T}},
\end{eqnarray}
where $\epsilon(-\mathbf{k},\omega)$ is the linear plasma dielectric function. Assuming a static form of the dielectric function, $\epsilon(-\mathbf{k},\omega)\approx1+k_{sc}^{2}/k^{2}$, where $k_{sc}$ is an inverse screening length, the Gaunt factor can be written
\begin{eqnarray}
\label{eq:GauntFactorSFCorrected1}
    G(\omega) \approx \frac{\sqrt{3}}{\pi}\int_{0}^{k_{\max}}\frac{dkk^{3}}{(k^{2}+k_{sc}^{2})^{2}}\exp{\left(-\frac{k_{\min}^{2}}{k^{2}}\right)}.
\end{eqnarray}
The integral is performed and then two limits are taken. The first is $k_{\min}/k_{\max}\ll1$, which is the same limit taken in the derivation of Oster's formulas. The second is $k_{sc}/k_{\max}\ll1$, which is well justified in a weakly coupled and non-degenerate plasma. Then the Gaunt factor can be written as
\begin{eqnarray}
\label{eq:GauntFactorSFCorrected2}
    \nonumber G(\omega) \approx \frac{\sqrt{3}}{2\pi}\left[\left(1-y^{2}\right)e^{y^{2}}\mathrm{Ei}\left(-y^{2}\right)-1\right]\\
    +\frac{\sqrt{3}}{\pi}\ln{\left(\frac{1}{e^{\gamma/2}}\frac{k_{\max}}{k_{\min}}\right)}
\end{eqnarray}
where $y\equiv k_{\min}/k_{sc}$ and $\mathrm{Ei}(x)$ is the exponential integral. The second term on the right hand side is just Oster's Gaunt factor. The first term is a correction to this that accounts for static screening. In the very low frequency limit $y\ll1$ and for $k_{sc} = 1/\lambda_{De}$ this returns the Dawson and Oberman results in Eqns.~(\ref{eq:dawsonandoberman}) and (\ref{eq:dawsonandobermanquant}).

\begin{figure}[h!]\label{screening_correction}
\includegraphics[width=0.5\textwidth]{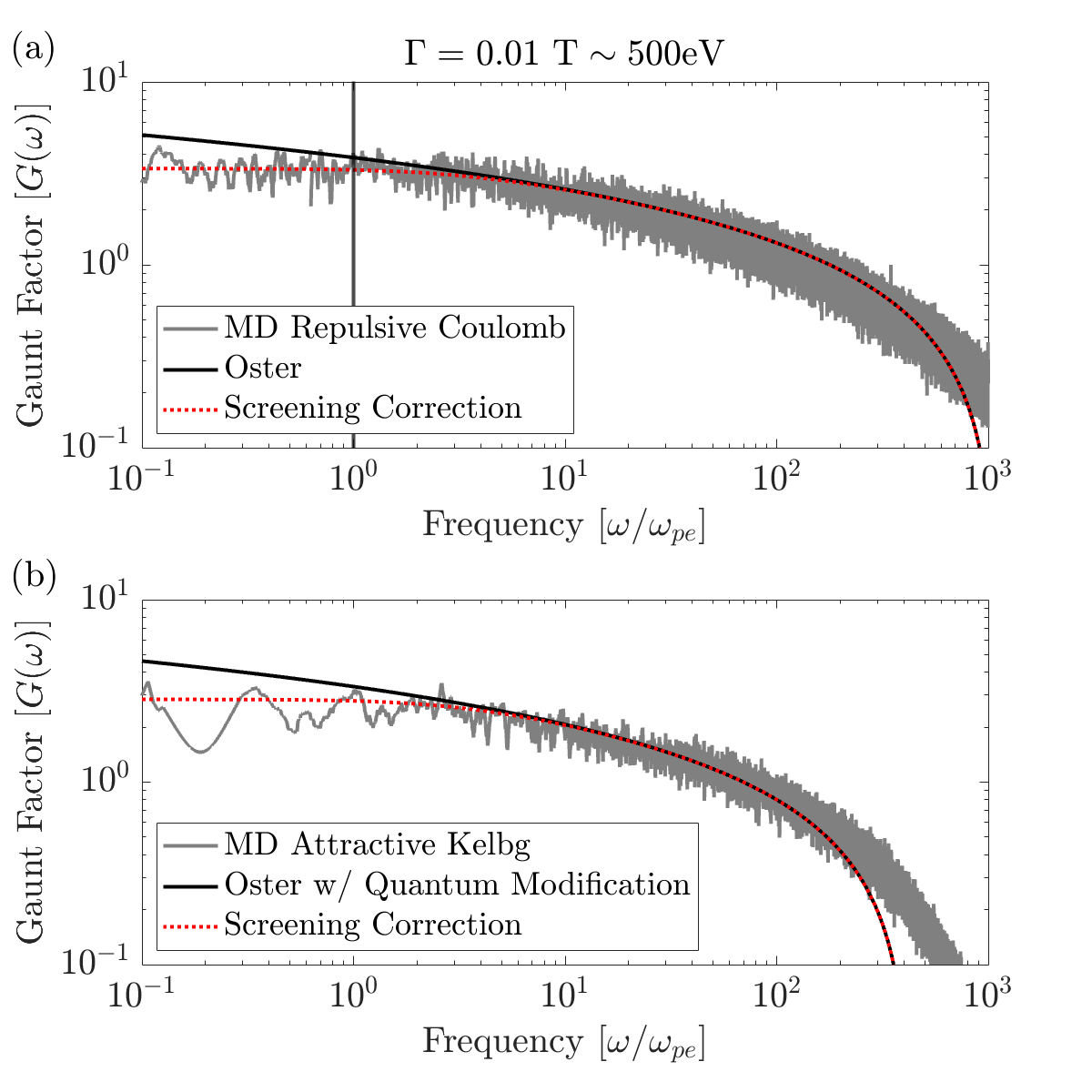}
\caption{\label{screening_correction} Comparisons of the bremsstrahlung Gaunt factor from MD simulations, Oster's formula, and the screening correction provided in Eq.~(\ref{eq:GauntFactorSFCorrected2}). The inverse screening length is set so that $k_{sc} = 1/\lambda_{D}$. (a) The MD simulations use the repulsive Coulomb potential, the Oster result uses Eq.~(\ref{eq:GauntFactorClassicalCoulombLowFreq}), and $k_{\max} = 4/(e^{2\gamma}r_{L})$. (b)The MD simulations use the attractive Kelbg potential, the Oster result uses Eq.~(\ref{eq:GauntFactorClassicalCoulombLowFreqQuantumMod}), and $k_{\max} = 2/(e^{\gamma/2}\lambda)$}
\end{figure}

Figure~\ref{screening_correction} shows that this correction agrees qualitatively with the plateau that occurs near the electron plasma frequency in the calculations from MD. The inverse screening length used in the plot is $k_{sc} = 1/\lambda_{D}$. It is noted that the practical evaluation of the product $e^{y^{2}}\mathrm{Ei}\left(-y^{2}\right)$ is prone to overflow errors. In our evaluation we employ the approximation given by Barry~\cite{BARRY2000287} where
\begin{eqnarray}
    e^{y^{2}}\mathrm{Ei}\left(-y^{2}\right) \approx -\frac{\ln{\left[1+\frac{G}{y^{2}}-\frac{1-G}{\left(h+by^{2}\right)^{2}}\right]}}{G+(1-G)\exp{\left(-\frac{y^{2}}{1-G}\right)}}
\end{eqnarray}
and
\begin{eqnarray}
    \nonumber G = \exp{\left(-\gamma\right)}\\
    \nonumber b = \sqrt{\frac{2(1-G)}{G(2-G)}}\\
    \nonumber h = \frac{1}{1+y^{3}} + \frac{h_{\infty}q}{1+q}\\
    \nonumber h_{\infty} = \frac{(1-G)(G^{2}-6G+12)}{3G(2-G)^{2}b}\\
    \nonumber q = \frac{20}{47}y^{\sqrt{62/13}}.
\end{eqnarray}

\bibliography{aipsamp}

\end{document}